\documentclass[epj]{svjour}
\usepackage{graphicx}
\usepackage{lineno}
\usepackage{amssymb}
\usepackage{units}
\usepackage{amsmath}
\usepackage{booktabs}
\usepackage{amsmath}
\usepackage{dblfloatfix}
\usepackage{color}
 
\newcommand{\kap}{K$^+$}
\newcommand{\kam}{K$^-$}
\newcommand{\pim}{$\pi^-$}
\newcommand{\pip}{$\pi^+$}

\begin{document}
\title{Strange meson production in Al+Al collisions at 1.9A~GeV}
\author{ 
P.~Gasik\inst{1,2,3} \and
K.~Piasecki\inst{1} \and
N.~Herrmann\inst{4} \and
Y.~Leifels\inst{5} \and
T.~Matulewicz\inst{1} \and
A.~Andronic\inst{5} \and
R.~Averbeck\inst{5} \and
V.~Barret\inst{6} \and
Z.~Basrak\inst{7} \and
N.~Bastid\inst{6} \and
M.L.~Benabderrahmane\inst{4} \and
M.~Berger\inst{3} \and
P.~Buehler\inst{8} \and
M.~Cargnelli\inst{8} \and
R.~\v{C}aplar\inst{7} \and
P.~Crochet\inst{6} \and
O.~Czerwiakowa\inst{1} \and
I.~Deppner\inst{4} \and
P.~Dupieux\inst{6} \and
M.~D\v{z}elalija\inst{9} \and
L.~Fabbietti\inst{2,3} \and
Z.~Fodor\inst{10} \and
I.~Ga\v{s}pari\'c\inst{7} \and
Y.~Grishkin\inst{11} \and
O.N.~Hartmann\inst{5} \and
K.D.~Hildenbrand\inst{5} \and
B.~Hong\inst{12} \and
T.I.~Kang\inst{5,12} \and
J.~Kecskemeti\inst{10} \and
Y.J.~Kim\inst{5} \and
M.~Kirejczyk\inst{1,13} \and
M.~Ki\v{s}\inst{5,7} \and
P.~Koczon\inst{5} \and
R.~Kotte\inst{14} \and
A.~Lebedev\inst{11} \and
A.~Le F\`{e}vre\inst{5} \and
J.L.~Liu\inst{15} \and
X.~Lopez\inst{6} \and
V.~Manko\inst{16} \and
J.~Marton\inst{8} \and
R.~M\"{u}nzer\inst{2,3} \and
M.~Petrovici\inst{17} \and
F.~Rami\inst{18} \and
A.~Reischl\inst{4} \and
W.~Reisdorf\inst{5} \and
M.S.~Ryu\inst{12} \and
P.~Schmidt\inst{8} \and
A.~Sch\"{u}ttauf\inst{5} \and
Z.~Seres\inst{10} \and
B.~Sikora\inst{1} \and
K.S.~Sim\inst{12} \and
V.~Simion\inst{17} \and
K.~Siwek-Wilczy\'{n}ska\inst{1} \and
V.~Smolyankin\inst{11} \and
K.~Suzuki\inst{8} \and
Z.~Tymi\'{n}ski\inst{4} \and
P.~Wagner\inst{18} \and 
I.~Weber\inst{9} \and
E.~Widmann\inst{8} \and
K.~Wi\'{s}niewski\inst{1,4} \and
Z.G.~Xiao\inst{19} \and
I.~Yushmanov\inst{16} \and
Y.~Zhang\inst{4,20} \and
A.~Zhilin\inst{11} \and
V.~Zinyuk\inst{4} \and
J.~Zmeskal\inst{8}
}                     
\offprints{}          
\institute{
Institute of Experimental Physics, Faculty of Physics, University of Warsaw, Warsaw, Poland
\and Excellence Cluster 'Origin and Structure of the Universe', Garching, Germany
\and Physik Department E62, Technische Universit\"{a}t M\"{u}nchen, Garching, Germany
\and Physikalisches Institut der Universit\"{a}t Heidelberg, Heidelberg, Germany
\and GSI Helmholtzzentrum f\"{u}r Schwerionenforschung GmbH, Darmstadt, Germany 
\and Laboratoire de Physique Corpusculaire, IN2P3/CNRS, and Universit\'{e} Blaise Pascal, Clermont-Ferrand, France 
\and Ru{d\llap{\raise 1.22ex\hbox{\vrule height 0.09ex width 0.2em}}\rlap{\raise 1.22ex\hbox{\vrule height 0.09ex width 0.06em}}}er Bo\v{s}kovi\'{c} Institute, Zagreb, Croatia
\and Stefan-Meyer-Institut f\"{u}r subatomare Physik, \"{O}sterreichische Akademie der Wissenschaften, Wien, Austria
\and University of Split, Split, Croatia
\and Wigner RCP, RMKI, Budapest, Hungary
\and Institute for Theoretical and Experimental Physics, Moscow, Russia
\and Korea University, Seoul, Korea
\and National Centre for Nuclear Research, Otwock-\'{S}wierk, Poland
\and Institut f\"{u}r Strahlenphysik, Helmholtz-Zentrum Dresden-Rossendorf, Dresden, Germany
\and Harbin Institute of Technology, Harbin, China
\and National Research Centre 'Kurchatov Institute', Moscow, Russia
\and Institute for Nuclear Physics and Engineering, Bucharest, Romania
\and Institut Pluridisciplinaire Hubert Curien and Universit\'{e} de Strasbourg, Strasbourg, France
\and Department of Physics, Tsinghua University, Beijing, China
\and Institute of Modern Physics, Chinese Academy of Sciences, Lanzhou, China
}
\date{Received: date / Revised version: date}
%
\abstract{
The production of \kap, \kam\ and 
$\phi$(1020) mesons is studied in Al+Al collisions at a beam 
energy of 1.9A~GeV which is close to or below the production threshold in NN
reactions. 
 Inverse slopes, anisotropy parameters, 
and total emission yields of K$^{\pm}$ mesons are obtained. 
A comparison of the ratio of kinetic energy distributions of 
\kam\ and \kap\ mesons to the HSD transport model calculations suggests that the inclusion of the in-medium modifications 
of kaon properties is necessary to reproduce the ratio.
The inverse slope and total yield of $\phi$ mesons are deduced.
The contribution to \kam\ production from $\phi$ meson decays is found to be
$\left[17 \pm 3\text{(stat)} ^{+2}_{-7}\text{(syst)}\right]\%$.
The results are in line with the previous K$^{\pm}$ and $\phi$ 
data obtained for different colliding systems at similar incident beam energies. 
} 
\maketitle
%

\section{Motivation}
\label{pga:motiv}
Strange particles are very sensitive probes of hot and dense nuclear 
matter formed in relativistic nucleus-nucleus collisions. 
It is predicted that the kaon-nucleon 
(KN) interaction is modified in dense nuclear matter 
with respect to the one in vacuum 
\cite{Brown91,Weise96,Waas97,Lutz04,Fuch06,Hart11}. Kaons (\kap\ and K$^0$) are
subject to a repulsive potential, whereas antikaons 
(\kam\ and $\overline{\text{K}}^0$) are attracted in a dense nuclear
medium. 
As a result of the in-medium KN interaction 
the effective mass of kaon, and the threshold energy for its production
should increase, whereas for antikaons
the corresponding values should decrease substantially \cite{Scha97}. 
The in-medium modifications of kaonic properties have been already studied
experimentally and reported by several experiments
focused on strangeness production  
at near-threshold energies \cite{Wisn00,Fors07,Bena09,Agak10,Ziny14}. 
Conclusions were based on the comparison to the results of theoretical
transport models.

\begin{figure*}
 \centering
  \begin{tabular}{lcr}
  \begin{minipage}[c]{0.3\textwidth}
  \includegraphics[width=6cm]{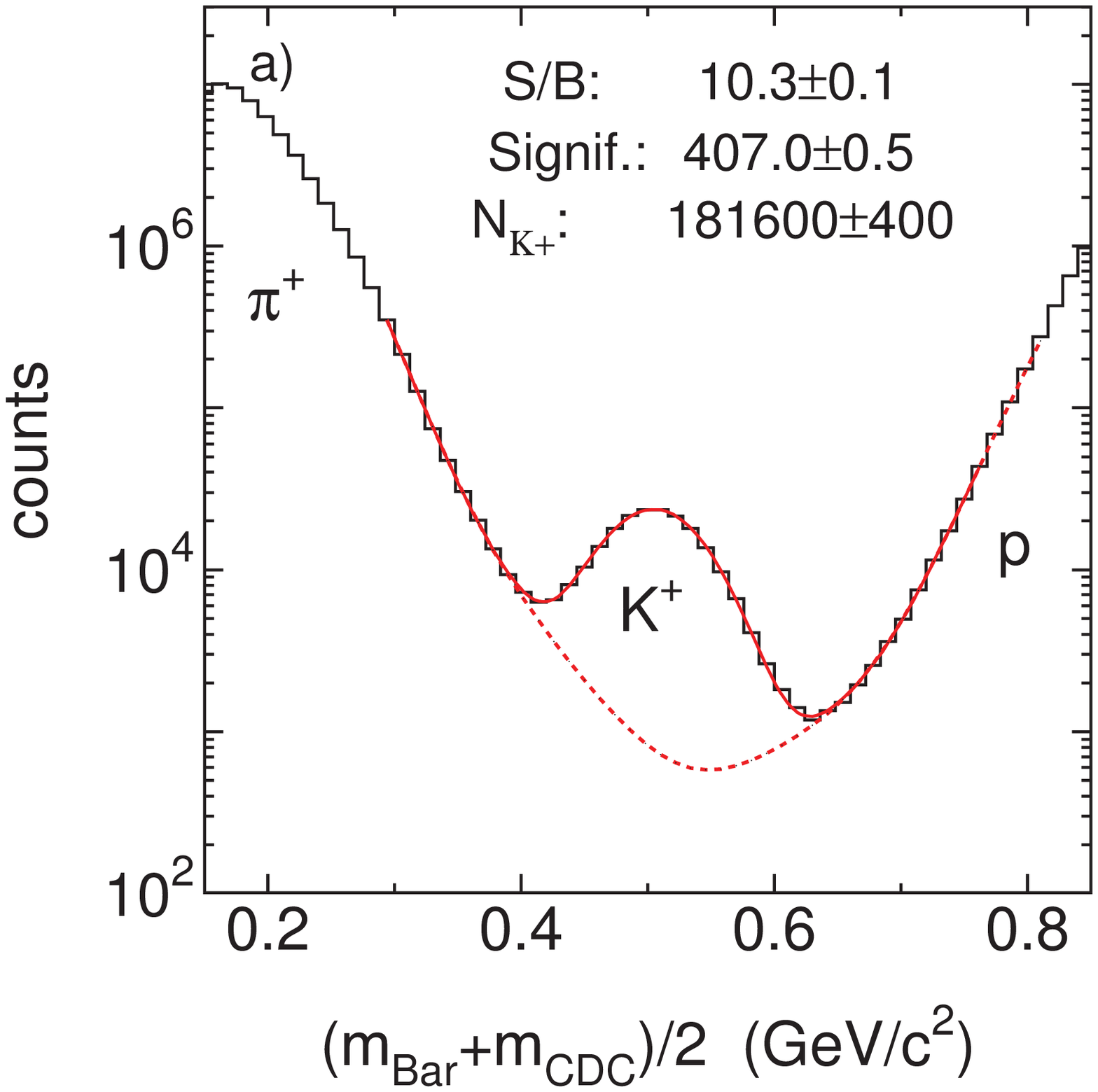}
  \end{minipage}
&
  \begin{minipage}[c]{0.3\textwidth}
  \includegraphics[width=6cm]{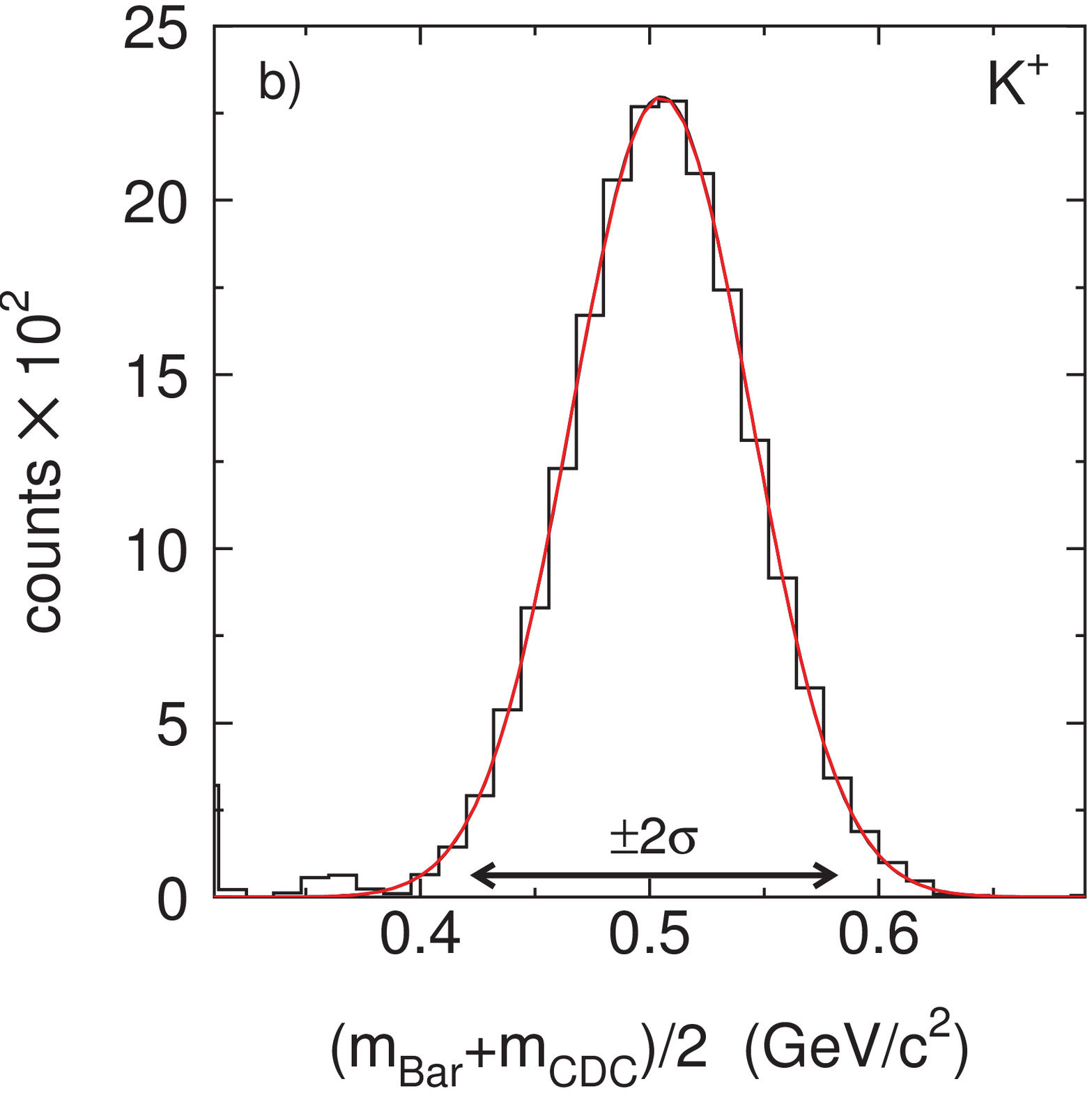}
  \end{minipage}
&
  \begin{minipage}[c]{0.3\textwidth}
  \includegraphics[width=6cm]{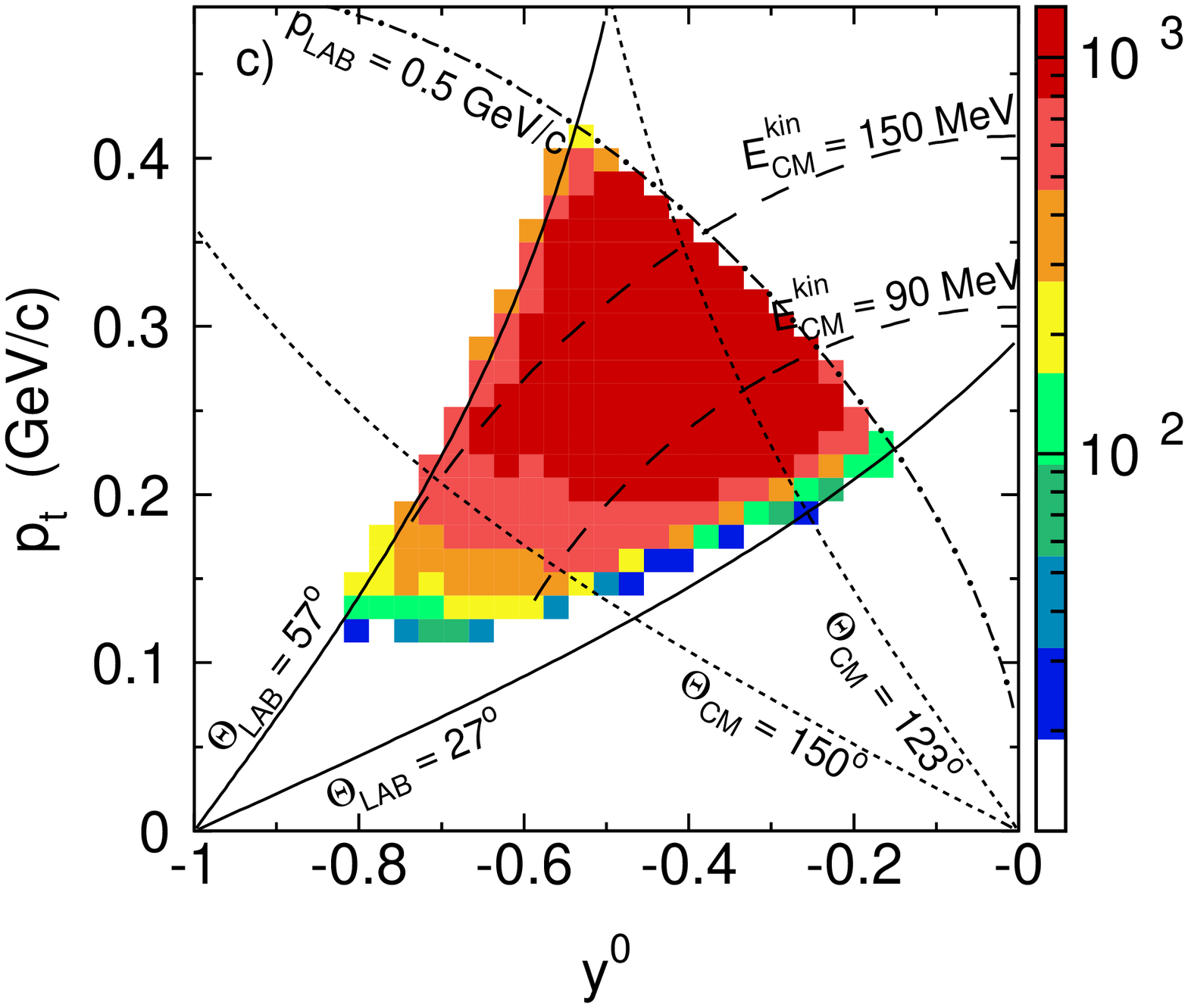}
  \end{minipage}
\end{tabular}

 \caption{ a) Experimental mass distribution of positively
  charged particles with momentum $p < 0.5$~GeV/c.
  The \kap\ signal is seen between the \pip\ and proton signals. 
  Solid (dotted) curve represent the signal (background) components
  of the function fitted to the distribution.
  b) \kap\ signal after background subtraction. The red curve represents  
  the Gaussian fit. 
  c) Phase space distribution of the reconstructed \kap\ mesons. 
  Solid and dashed-dotted lines depict the angular and momentum
  limits of the detector, respectively. Dashed (dotted) lines 
  represent the constant values of kinetic energy (polar angle)
  in the center-of-mass frame. See text for details.} 
 \label{fig:kp}
\end{figure*}

For decisive conclusions on in-medium modifications, one has to account 
for feeding of the \kam\ emission  by the $\phi$(1020) meson decays, since the 
$\phi$(1020) is decaying into a \kap\kam\ pair with a probability of 48.9~\%.
A comparison of the mean free path of $\phi$(1020) mesons, 
$c\tau_{\phi}\approx50$ fm, to the average duration of the 
hot and dense collision zone (20--30~fm/c~\cite{Hart11}) suggests
that most of $\phi$ mesons decay outside this zone. 
It should be also noted that the $\phi$ meson mass is only 
32~MeV/$c^2$ larger than that of a \kap\kam\ pair. 
As a result, \kam\ originating from
the $\phi$ meson decays are governed by different kinematics than 
those emitted directly from the collision zone. It is therefore
crucial to find out in the first step the relative contributions 
of these sources to the total \kam\ production. 

It has been reported for two systems, Ni+Ni and Ar+KCl, colliding at beam kinetic energies of 1.7--1.9A~GeV, that about 20\% of the \kam\ mesons originate  
from $\phi$ meson decays~\cite{Agak09,Lore10,Pias15}. 
Here, we present a third data point on this subject, supporting the
statement that $\phi$ meson decays are a relevant source of the 
\kam\ mesons in heavy-ion collision at energies below the NN production
threshold.  
We also propose a simple two-source model that aims to reconstruct 
the kinematic properties of direct kaons from the inclusive 
\kam\ spectra. 

\begin{figure*}[tbp]
 \centering
 \begin{tabular}{lcr}
  \begin{minipage}[c]{0.3\textwidth}
  \includegraphics[width=6cm]{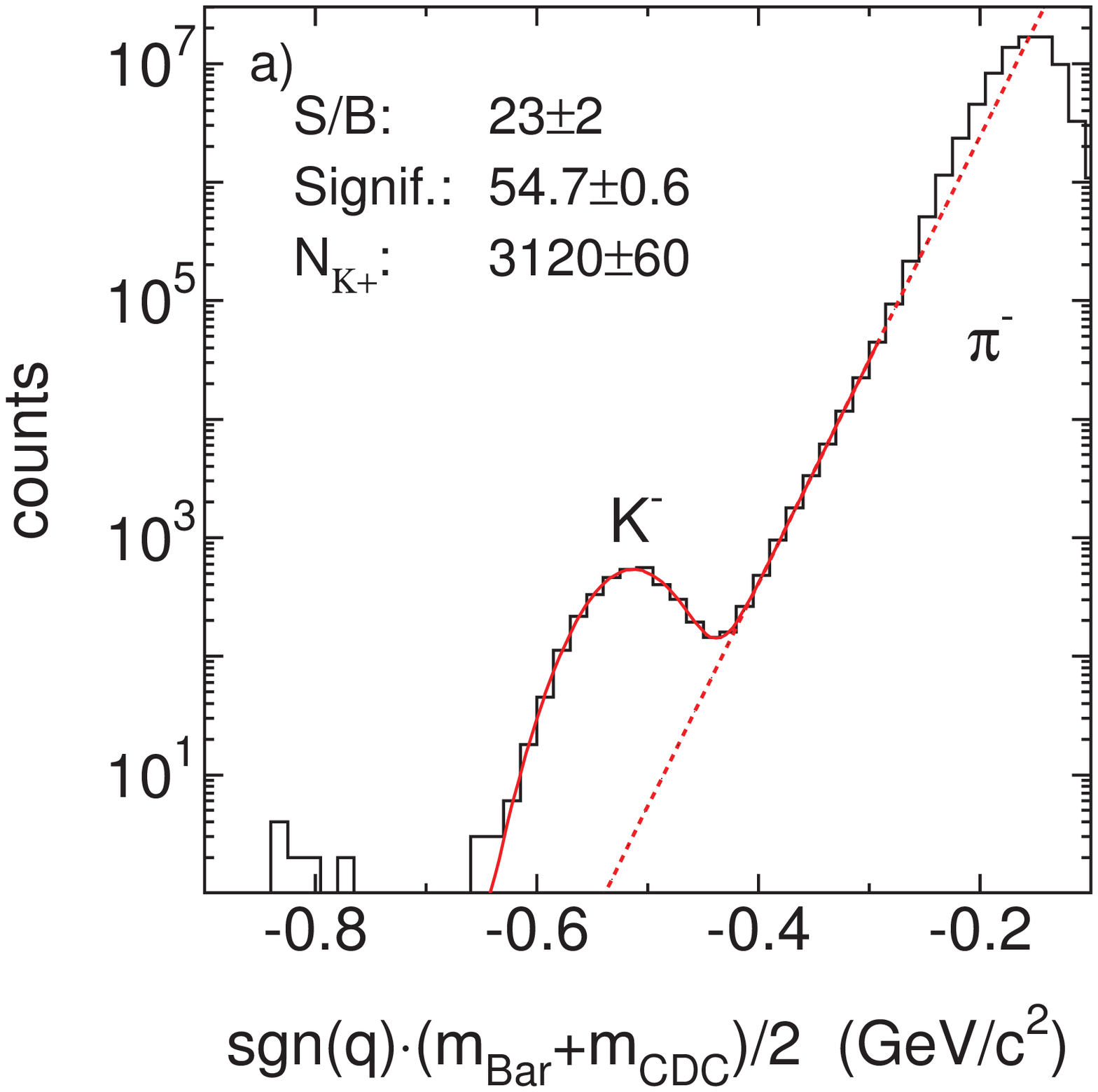}
  \end{minipage}
&
  \begin{minipage}[c]{0.3\textwidth}
  \includegraphics[width=6cm]{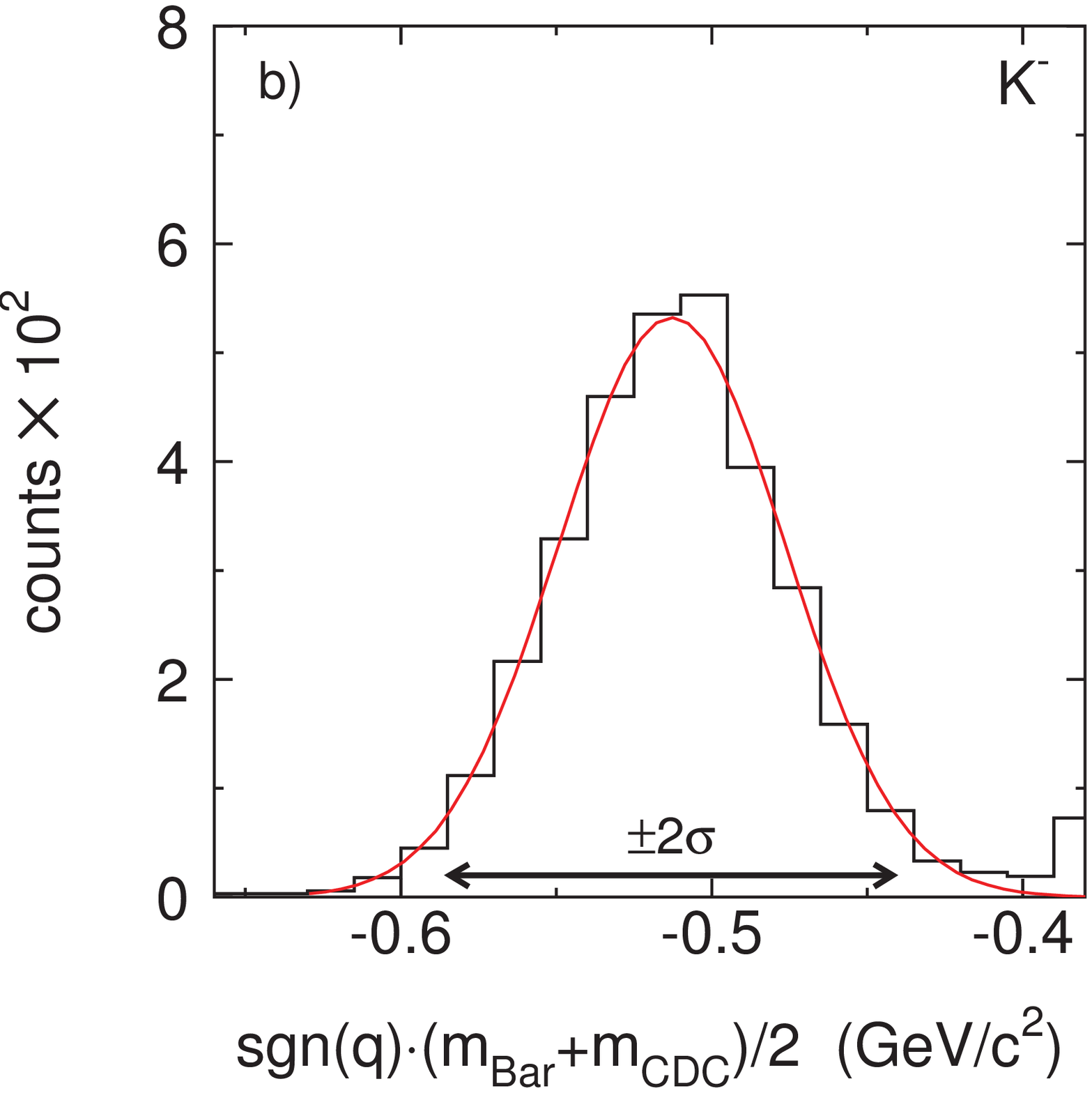}
  \end{minipage}
&
  \begin{minipage}[c]{0.3\textwidth}
  \includegraphics[width=6cm]{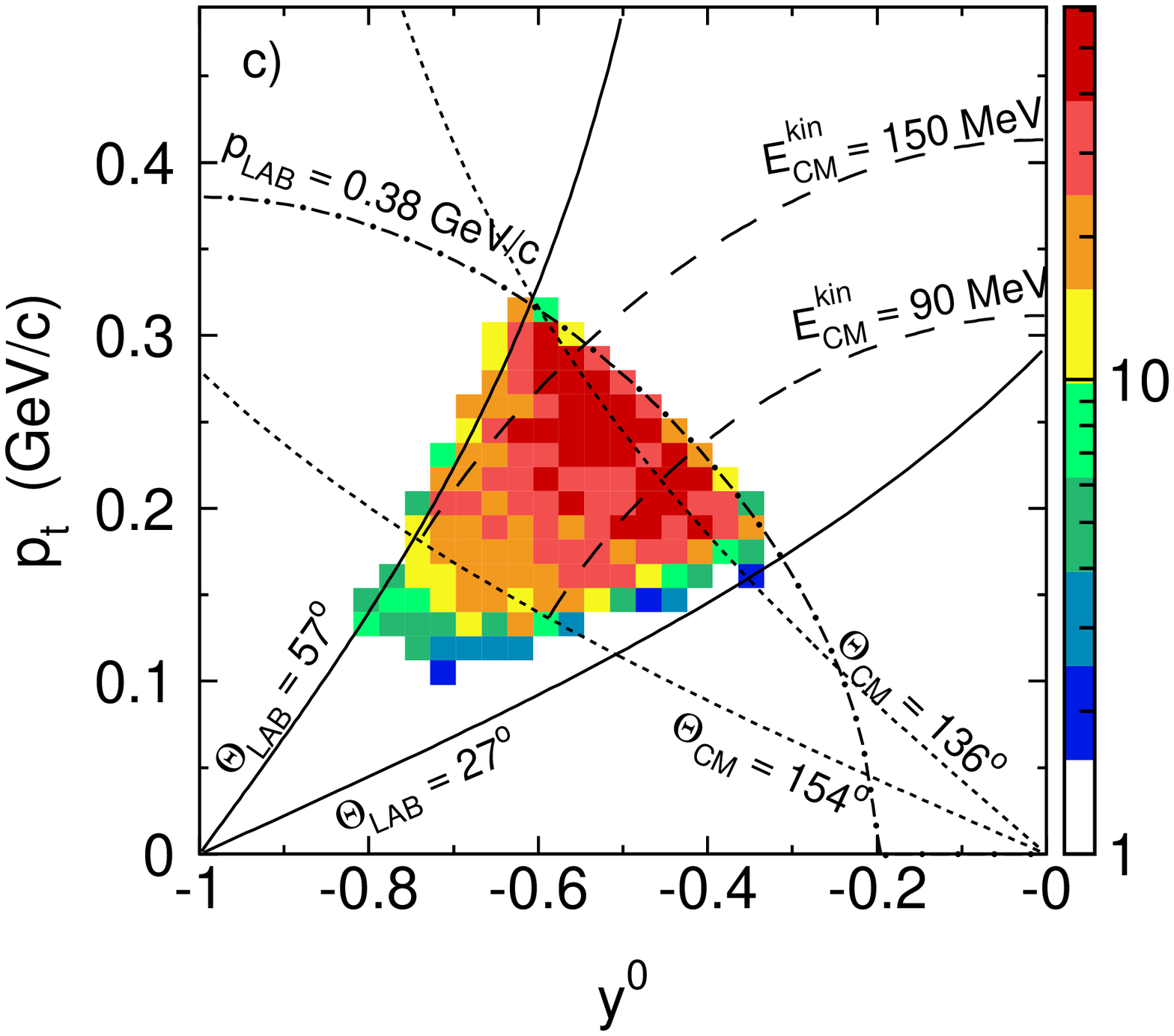}
  \end{minipage}
  \end{tabular}
 \caption{a) Experimental mass distribution of negatively 
  charged particles with momentum $p < 0.38$~GeV/$c$. 
  The \kam\ signal is seen on the tail of the pion signal. 
  Solid (dotted) curve represent the signal (background) components
  of the function fitted to the distribution.
  b) \kam\ signal after the background subtraction. The red curve
  represents the Gaussian fit.
  c) Phase space distribution of the reconstructed \kam\ mesons. 
  Solid and dashed-dotted lines depict the angular and momentum
  limits of the detector, respectively. Dashed (dotted) lines 
  correspond to the constant values of kinetic energy (polar angle) 
  in the center-of-mass frame. See text for details.}  
 \label{fig:km}
\end{figure*}

\section{FOPI spectrometer}
\label{pga:setup}
FOPI is a modular spectrometer for fixed-target experiments 
at the SIS-18 synchrotron in the GSI, Darmstadt. 
The FOPI set-up consists out of 4 subdetectors. 
Two drift chambers are mounted around the target, 
CDC (Central Drift Chamber) and Helitron. 
The CDC detector spans a wide range of polar angles
$27^\circ < \vartheta_\text{lab} < 113^\circ$.
It is surrounded by a Plastic Scintillation Barrel, 
a time-of-flight (ToF) detector covering polar angles
$27^\circ < \vartheta_\text{lab} < 57^\circ$.
These detectors are located inside the magnet solenoid with a magnetic field
of 0.6~T. 
The PlaWa (Plastic Scintillation Wall) detector is installed at the forward
directions behind the Helitron, outside the magnetic field. The target is
shifted 40~cm upstream with 
respect to the nominal position.

A particle moving inside the magnetic solenoid follows the path of a helix. 
If it passes through the CDC, electric signals are induced on the sense
wires (dubbed below "hits") which are digitized by a Flash-ADC system. 
A fast on-line algorithm extracts the characteristics of the signals 
(time with respect to the event start, amplitude, and length). 
An off-line procedure uses this information to obtain the drift time, energy loss, and position of a hit. 
Particle tracks are reconstructed by employing
the tracking algorithm based on a local approach searching for consecutive 
hits on a circle (straight line) in the plane transverse (longitudinal) 
to the beam axis.

Particle identification in the CDC is obtained by correlating 
the energy loss d$E/$d$x$ and the curvature of a track. 
Particle masses extracted from this procedure
shall be dubbed $m_{\text{CDC}}$. 
Plastic detectors, together with the Start counter placed about 2 meters
in front of the target, yield an additional ToF measurement of a particle.
For a CDC track, which can be matched with a hit in the ToF Barrel (or PlaWa),
the particle type is identified by employing the relativistic relation 
between momentum and velocity, $p = m\beta\gamma$, where $m$ is the mass
of a particle, $\beta$ its reduced velocity $v/c$, and $\gamma$ is the Lorentz
factor. The mass obtained by this procedure 
shall be dubbed $m_{\text{Bar}}$. The timing resolution of the plastic detector 
is crucial and limits the extension of the momentum space, for which 
the charged kaons can be identified.
More details on the configuration of FOPI submodules and their 
performance can be found in refs.~\cite{FOPI1,FOPI2}.

\begin{figure*}[tbp]
 \centering
 \includegraphics[width=14.cm]{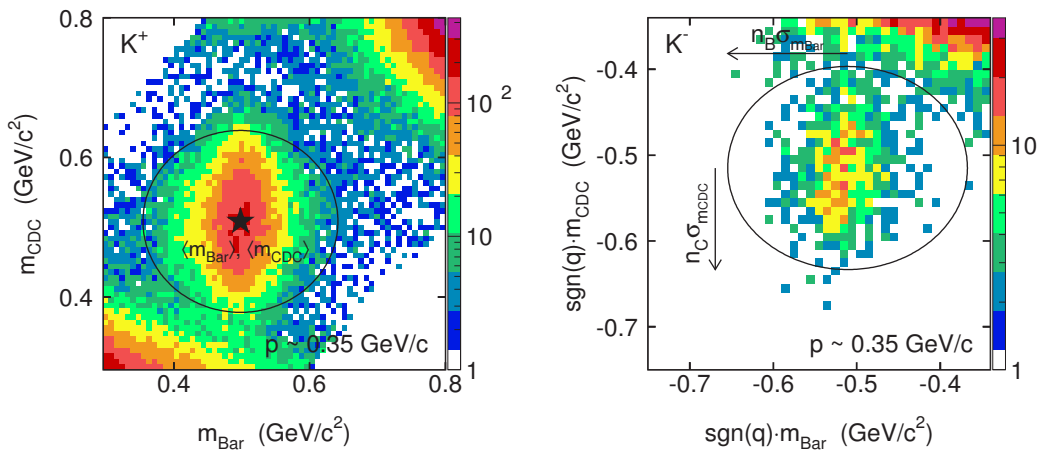}
 \caption[]{Correlation plot of $m_{\text{Bar}}$--$m_{\text{CDC}}$ mass distributions 
   of particles with momentum $p \approx 0.35~\text{GeV}/c^2$, 
   shown around the nominal masses of \kap\ (left panel) and \kam\ (right panel) mesons.
   Ellipses represent the cuts imposed for the $\phi$ meson identification.}
 \label{fig:fikmasscut}
\end{figure*}

\section{Data analysis}
\label{pga:ana}
\subsection{Al+Al experiment}

The aluminium target of 567\,mg/cm$^2$ thickness was irradiated by the beam of 
aluminium ions with an intensity of $8 \times 10^5 ~\text{s}^{-1}$ at the kinetic 
energy of 1.9A~GeV. 
The collision centrality was determined by the multiplicity of charged 
particles in the ToF Barrel and PlaWa detectors. A minimum multiplicity in the
Plastic Wall detector was required as a trigger condition to select 
the more central events. The number of collected events is $1.59 \times 10^8$.
The results reported in this paper are obtained for the most central 9\% 
of the total geometrical cross section ($\Delta\sigma\approx 140$~mb). 
Using a geometrical \textit{sharp cut-off} model, one can estimate 
the maximal value of impact parameter $b$ for the analysed data sample 
to be $\sim$2.1\,fm.
The mean number of participant nucleons was calculated using 
the \textit{participant-spectator} model described in~\cite{Goss77},
and is found to be $\langle A_\text{part} \rangle_\text{b} \approx 42$.

\begin{figure*}[tbp]
 \centering
  \begin{tabular}{cc}
  \begin{minipage}[c]{0.5\textwidth}
  \includegraphics[width=7.5cm]{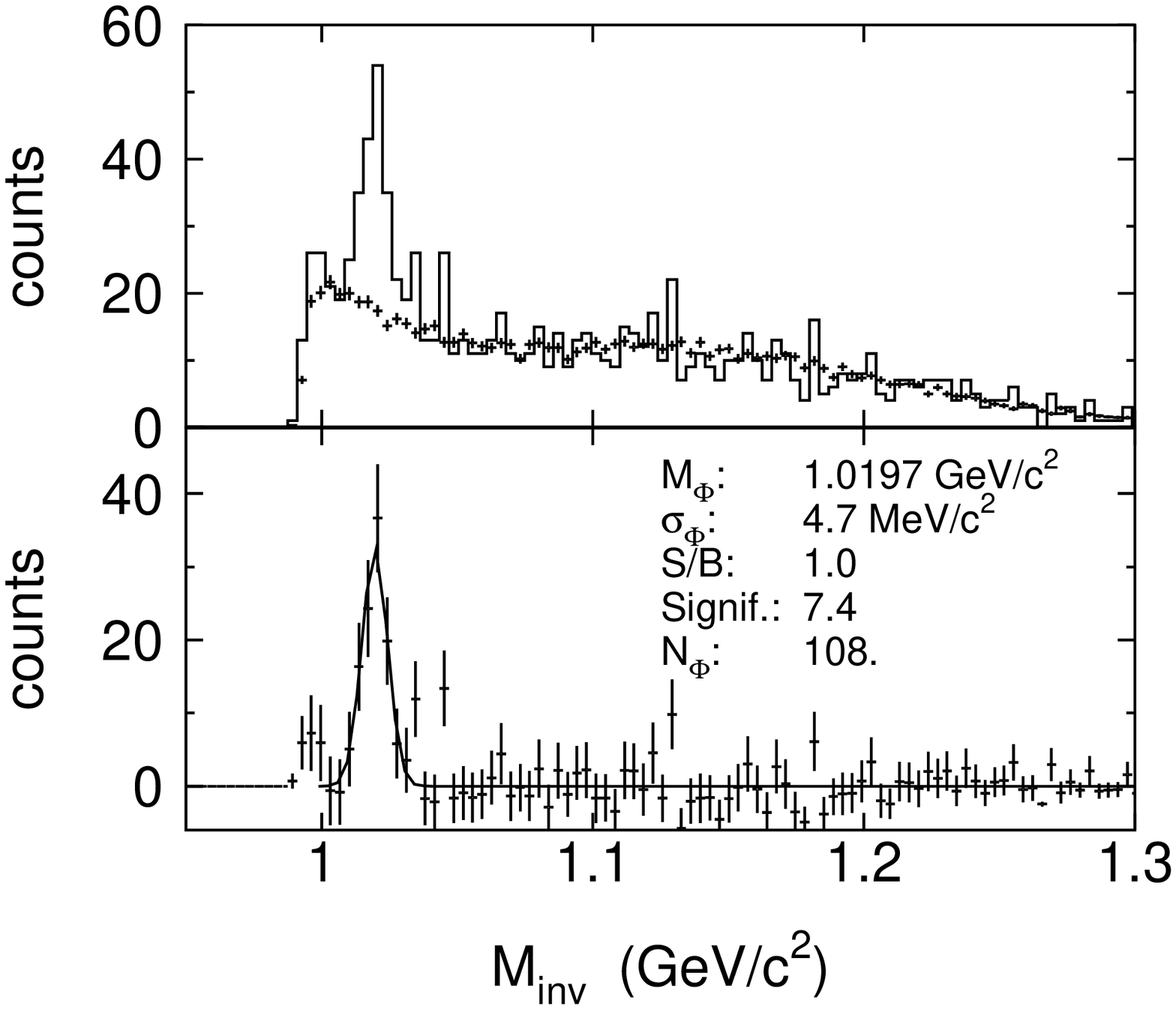}
  \end{minipage}
&
  \begin{minipage}[c]{0.5\textwidth}
  \includegraphics[width=7.5cm]{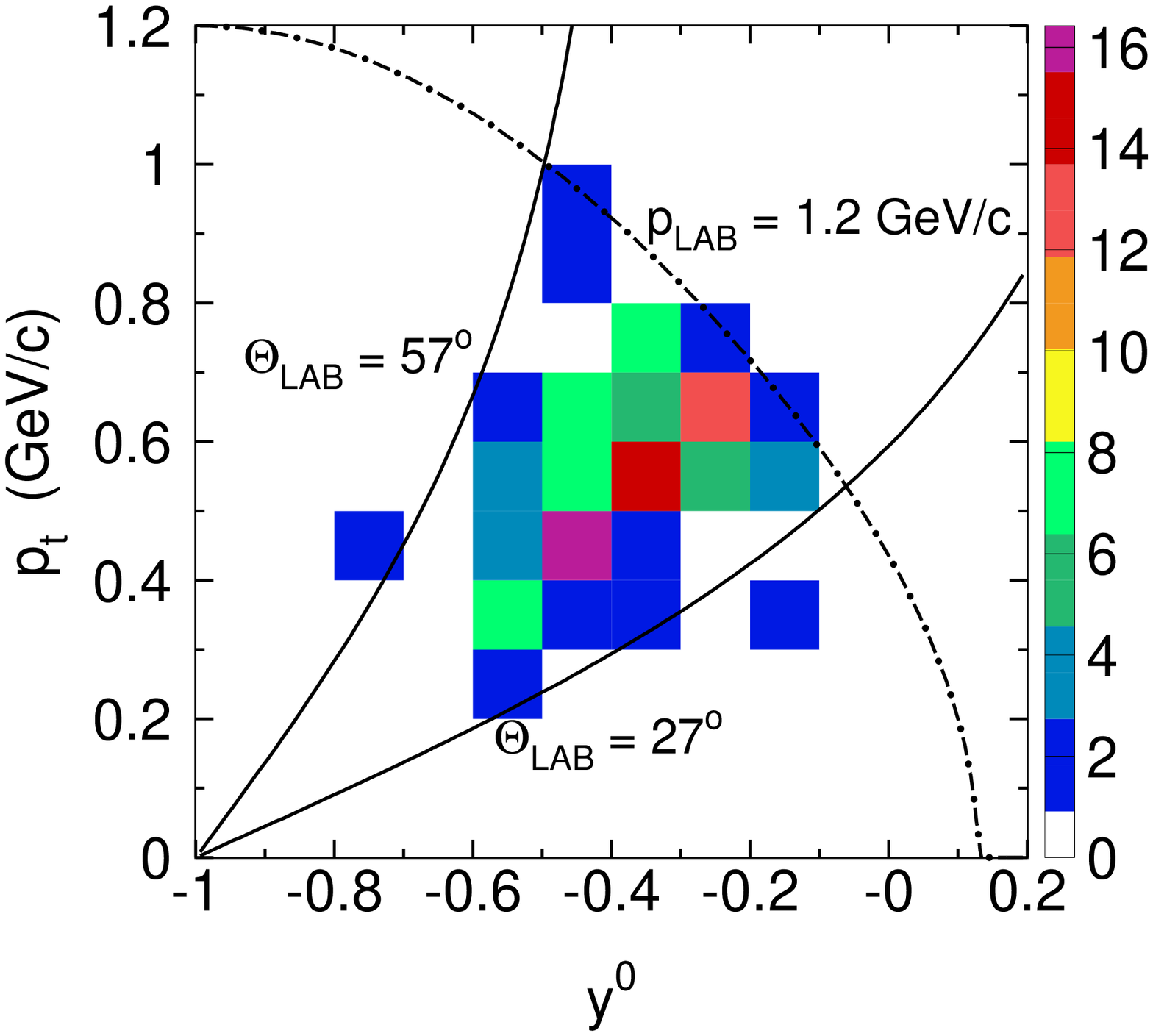}
  \end{minipage}
  \end{tabular}
 
 \caption[]{Left, upper panel: Invariant mass distribution of \kap\kam\ pairs. 
     The background spectrum of uncorrelated pairs is shown by the crosses. 
     Left, lower panel: The $\phi$ meson signal after the background subtraction, 
     fitted with a Gaussian function (solid curve). Right: $p_\text{t}-y^0$  
     phase space distribution of identified $\phi$ mesons. Solid lines depict the angular
     limits of the detector and the dashed-dotted line - the momentum cut.}
 \label{fig:minvfin}

\end{figure*}

\subsection{Kaon identification}
\label{sec:kaonid}

\kap\ and \kam\ mesons were identified by the CDC tracks 
matched with the hits in the ToF Barrel. 
Figs.~\ref{fig:kp}a and \ref{fig:km}a present the experimental mass 
distributions around $m_{\text{K}^{\pm}}$ of positively (negatively) 
charged particles with momenta $p < 0.5 ~(0.38)$~GeV/$c$, respectively. 

The \kap\ signal, concentrated around 0.5~GeV/$c^2$, is situated on the tails 
of the proton and \pip\ signals, which can be described by a combination 
of Gaussian and exponential functions (dotted line). The background under the 
\kam\ peak is caused by the \pim\ mesons, and can be described by an
exponential function. 

In order to enhance the signal-to-background ratio, and the significance 
of the signal, the following set of cuts was applied:
\begin{itemize}
 \item a limit on the \kap\ (\kam) momentum to \linebreak
      $p_\text{lab} < 0.5 ~(0.38)$~GeV/$c$,
       to reduce the contamination from pions and protons;
 \item a cut on the transverse momentum ($p_\text{t} > 0.1$~GeV/$c$), 
       to reject particles spiralling in the CDC;
 \item cuts on the distance of closest approach between the track and the
       collision vertex in the transverse plane ($d_0 < 1$~cm) 
       and its uncertainty ($\sigma_{\text{d}_0} < 0.14$~cm), 
       to suppress decay products outside the target;
 \item a condition on the minimum number of hits on a track
       in the CDC (35 for \kap\ and 40 for \kam) 
       for the higher quality of the reconstructed tracks;
 \item a cut on the angular difference between the extrapolation of the CDC
       track and the matched ToF Barrel hit, $\Delta\phi < 5^\circ$.
\end{itemize}
After background subtraction, clear peaks around the nominal K$^\pm$ mass 
are seen in figs.~\ref{fig:kp}b and \ref{fig:km}b. The solid
curves represent the Gaussian fits. Within a $\pm 2 \sigma$ region 
around the maximum of the fitted function (as indicated in the figs.~\ref{fig:kp}b 
and ~\ref{fig:km}b) about $180 \times 10^3$ \kap\ and $3 \times 10^3$ \kam\ 
mesons were reconstructed in the analysed data sample with a
signal-to-background ratio $S/B$ of 10 (\kap) and 23 (\kam), and a
significance (defined as $S/\sqrt{S+B}$) of 407 and 55 for \kap, and
\kam\ mesons respectively. 
The acceptance ranges for \kap, and \kam\ mesons are shown in
figs.~\ref{fig:kp}c and 
\ref{fig:km}c in terms of transverse momentum $p_{\text{t}}$, and normalised 
rapidity $y^0$, where $y^0 = y_{\text{lab}} / y_{\text{NN}} - 1$ is defined to 
be +1 (-1) at projectile (target) rapidity, and $y_{\text{NN}} \approx 0.89$.

The systematic uncertainties of the integrated background were estimated
by varying the function fit parameters. They are found to be at the level of 40\% 
(\kap) and 20\% 
(\kam). Note, that the corresponding signals are strong
enough for the influence of the above-mentioned errors to be minor. 

In order to estimate the systematic uncertainties associated with the cutting
strategy, apart from the "standard" set of cuts, "narrow" and "wide" cut
sets were applied to the parameters used for enhancing the quality of the
tracks and the matching of CDC with Plastic Barrel hits.
In case of the ''narrow'' cuts the background is minimised but at the same 
time the statistics  
of identified particles is reduced. The ''wide'' cuts are less restrictive. 
This results in an increase of the background but does not lead to a signal
increase at the same level.

In all the sets of cuts used in the charged kaons analysis, 
the total and transverse momentum cut were kept constant.

\subsection{$\phi$(1020) identification}

\begin{figure*}
 \centering
  \begin{tabular}{cc}
  \begin{minipage}[c]{0.5\textwidth}
  \includegraphics[width=9cm]{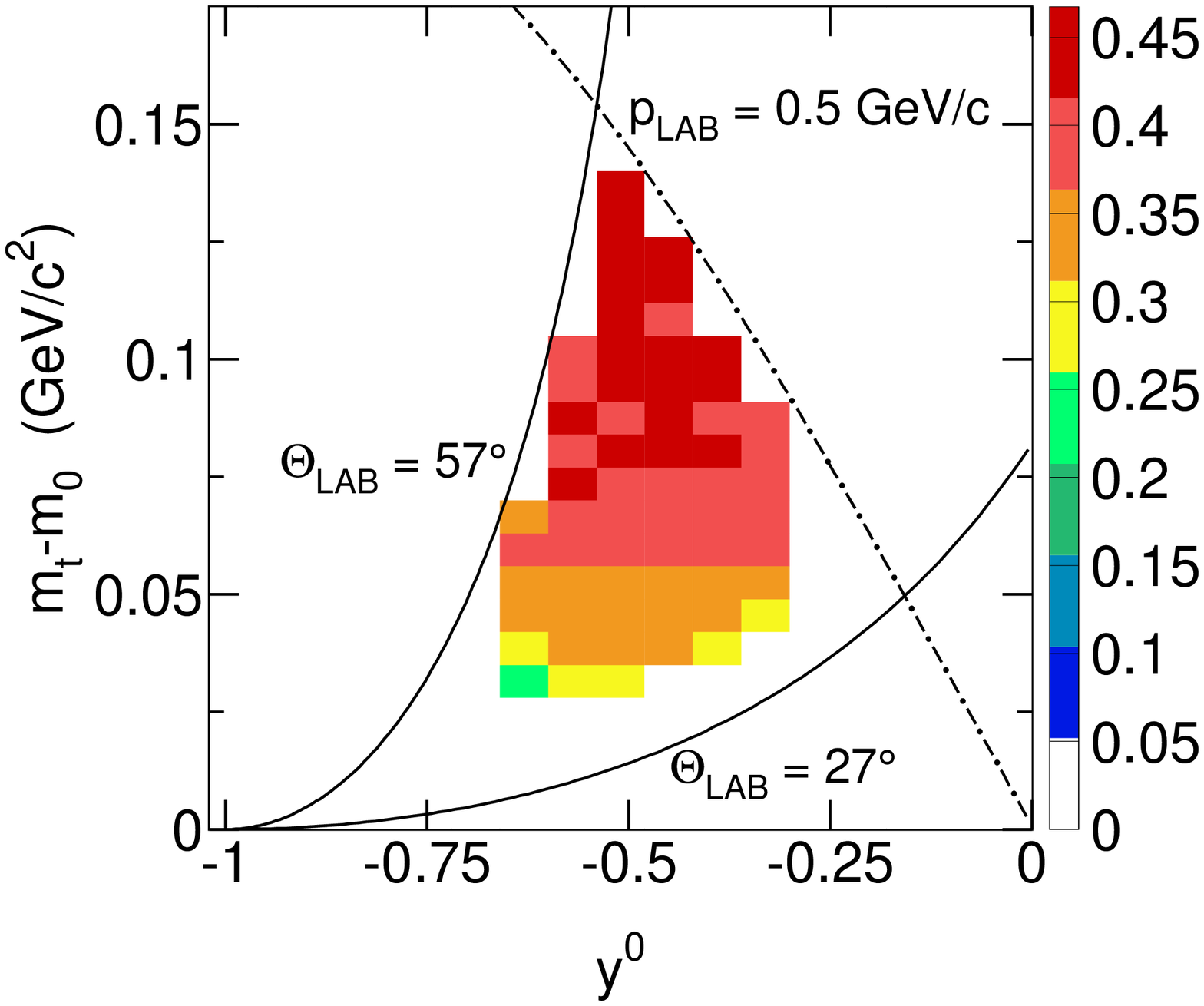}
  \end{minipage}
&
  \begin{minipage}[c]{0.5\textwidth}
  \includegraphics[width=9cm]{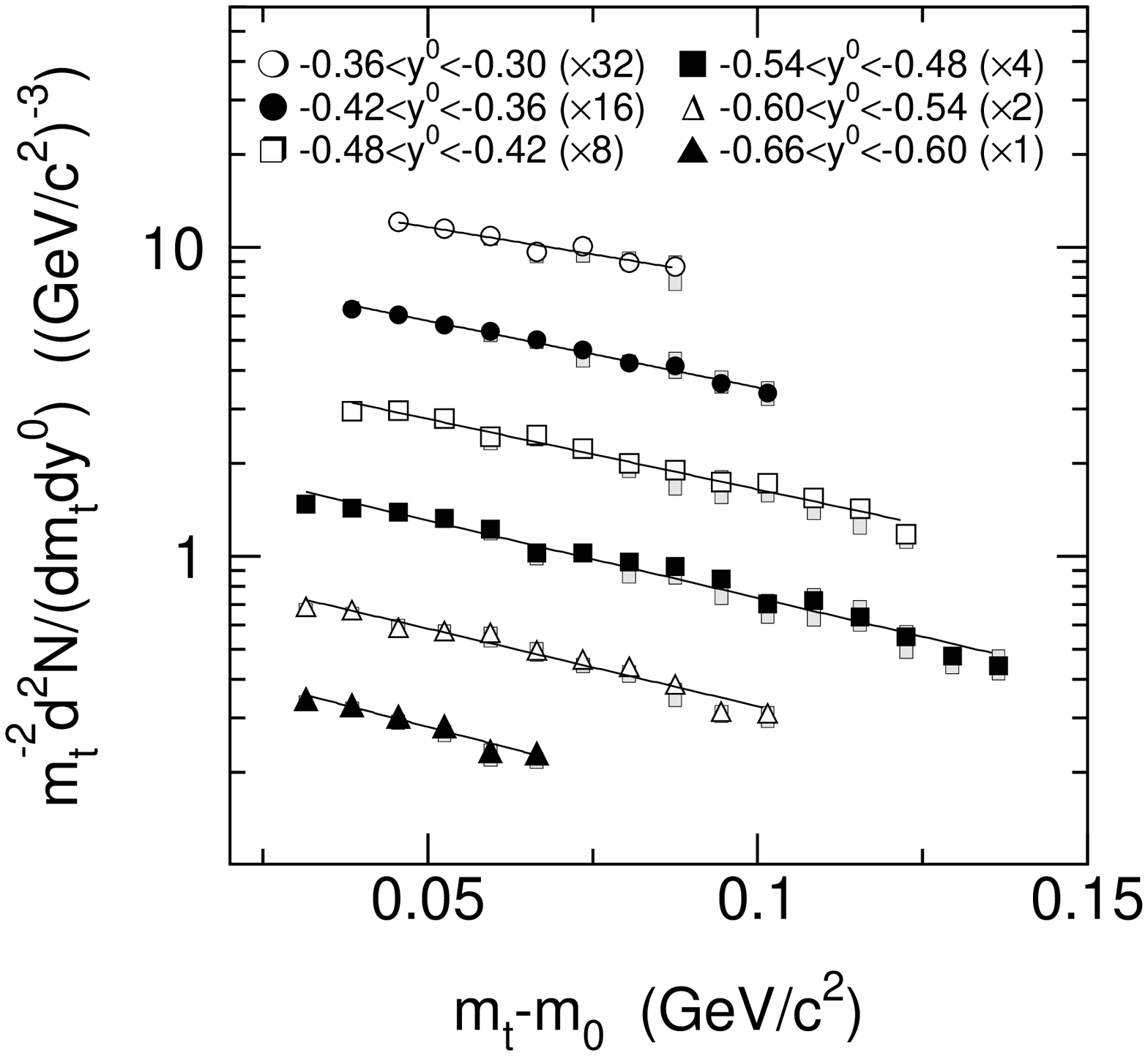}
  \end{minipage}
  \end{tabular}
  \caption{\label{fig:kpmt}Left: efficiency of the \kap\ meson identification 
   for the different bins of $m_\text{t} - m_0$ and $y^0$ used for the transverse 
   mass spectra analysis (see sec.~\ref{sec:keff} for more details). 
   Right: transverse mass spectra of \kap\ mesons. The spectra 
   measured in different slices of normalised rapidity are multiplied by the 
   subsequent powers of two. Solid lines indicate the Boltzmann function fits. 
   Shaded rectangles correspond to the systematic uncertainties.}
\end{figure*}

Identification of $\phi$ mesons is possible 
via their charged decay in the FOPI spectrometer: $\phi\rightarrow
\text{K}^+\text{K}^-$.  
The branching ratio for this channel is 
$\Gamma_{\text{K}^ + \text{K}^-}/\Gamma = 48.9 \pm 0.5\,\%$~\cite{PDG}.
Hence, $\phi$ mesons can be identified by an invariant mass analysis of 
\kap\kam\ pairs. Due to the short lifetime of $\phi$ mesons 
($c\tau \approx 1.55 \times 10^{-22}$\,s) it is not possible to disentangle 
its decay vertex from the collision vertex. 

In order to increase the identified $\phi$ meson signal, the maximum momentum
conditions were increased for this analysis to $p_{\text{lab}}
<0.6$\,GeV/$c$. However, such an increase may result in an inclusion of
some non-kaons ($\pi^{\pm}$, protons) in the sample of kaon pairs. 
To reduce such a side effect, a geometrical (2D-ellipse) cut on the mass of 
\kap\ and \kam\ mesons was applied, as shown in fig.~\ref{fig:fikmasscut}.

For the further $\phi$ analysis, only the kaons fulfilling the following 
ellipse equations were chosen:

\begin{equation}
 \left(\frac{m_{\text{Bar}}-\langle
   m_{\text{Bar}}\rangle}{n_\text{B}\sigma_{m_{\text{Bar}}}}\right)^2 +  
 \left(\frac{m_{\text{CDC}}-\langle
   m_{\text{CDC}}\rangle}{n_\text{C}\sigma_{m_{\text{CDC}}}}\right)^2 < 1\quad
 , 
\end{equation} 

\noindent where $m_{\text{Bar}}$, $m_{\text{CDC}}$ are the mass parameters 
(see sec.~\ref{pga:setup}), 
$n_\text{B}$, $n_\text{C}$ are the factors defining widths of the 
mass cuts for kaons, 
$\langle m_{\text{Bar}}\rangle$, $\langle m_{\text{CDC}}\rangle$ are the mass
parameter mean values, and 
$\sigma_{m_{\text{Bar}}}$, $\sigma_{m_{\text{CDC}}}$ are the dispersions of the 
kaon mass distributions.
The values of four latter parameters are momentum- and charge-dependent. 
The widths of ellipses given by $n_\text{B} = 4.5$ and $n_\text{C} = 2$, 
were chosen for the further analysis. 
An influence of the cut on the systematic uncertainty of the 
$\phi$ meson yield was estimated by varying these values.

Fig.~\ref{fig:minvfin} shows the invariant mass plot of $\phi$ mesons 
reconstructed in the CDC+Barrel subsystem with a clear indication of a peak
close to the nominal $\phi$(1020) mass. 

The combinatorial background was reconstructed using the event-mixing method, 
where matched kaons were selected from different events. In order to bring 
the kinematic conditions of mixed pairs close to those of true ones, 
two conditions were imposed:
\begin{itemize}
 \item \kap\ were selected from events exhibiting at least one \kam,
 \item The CDC hit multiplicity distribution, correlated with the collision
       centrality, was divided into 8 groups (centrality classes). Kaons were
       matched from events attributed to the same centrality class. 
\end{itemize}

The background distribution was adjusted to the one of true pairs by normalising
the spectrum to the latter one in the region $M_\text{inv} > 1.05$~GeV/$c$. 
The resulting combinatorial background spectrum is indicated with  
crosses in fig.~\ref{fig:minvfin} (upper left panel). 
108 $\phi$ mesons were reconstructed within $\pm 2\sigma$ range around 
the centre of the fitted Gaussian function, which is shown as full line in
fig.~\ref{fig:minvfin} (lower left panel). The distribution of reconstructed
$\phi$ mesons in the $p_{\text{t}}$ versus $y^0$ plane is shown in the 
right panel of fig.~\ref{fig:minvfin}.
In order to extract the $\phi$ meson yield, the efficiency of the 
detector was deduced using the GEANT package (see sec.~\ref{sec:fieff}).

\subsection{Efficiency correction}

Simulations of the detection efficiency of K$^\pm$ and $\phi$ mesons were
performed in the framework of the GEANT~\cite{GEANT} environment. 
Single \kap, \kam\ or $\phi$ mesons were added to events of Al+Al collisions
generated by the IQMD model~\cite{IQMD}, 
which served as a realistic background for tracking and PID. 
All particles of the combined event were tracked and digitised 
within GEANT, and subsequently reconstructed by the same off-line 
algorithms as that used for the experimental data.

\subsubsection{\kap\ and \kam\ mesons}
\label{sec:keff}

Charged kaons were generated from a uniform $p_\text{t} - y^0$ 
distribution, and each K$^\pm$ meson was subsequently added to the collection
of particles in an event generated by the IQMD code. 
In total, $~6.5\times10^6$ events were simulated.
For each studied kinematical distribution, which would be used further in
sec.~\ref{pga:kres}), a corresponding efficiency map was evaluated. 
It allowed for correcting these spectra on a bin-by-bin basis. 

As an example, the distribution of the \kap\ meson efficiency
as a function of transverse mass $m_\text{t}$ and normalised rapidity $y^0$ 
is shown in the left panel of fig.~\ref{fig:kpmt}. The covered phase space 
region reflects the acceptance of the CDC+Barrel detectors. 

To check whether the choice of the primary distribution of charged kaons
influences the efficiency map and thus the final results, K$^\pm$ mesons
were also sampled from a Boltzmann-like distribution multiplied by
the angular anisotropy term:

\begin{equation}
  \begin{aligned}
  \label{eq:EkThe}
  \frac{\text{d}^2 N}{\text{d}E_\text{CM}^\text{kin} \text{dcos}\vartheta_\text{CM}} = &
  N_0 \cdot p_\text{CM}E_\text{CM} \exp (-E_\text{CM}/T_\text{eff}) \cdot\\
  	&  (1 + a_2 \cos^2 \vartheta_\text{CM} )\quad , 
  \end{aligned}
\end{equation}

\noindent where $N_0$ is the overall normalisation factor, 
$E_\text{CM}$ and $p_\text{CM}$ are 
the energy and momentum of a particle in the center-of-mass frame,
$T_\text{eff}$ is the inverse slope parameter 
("effective temperature" of a particle source), and  
$a_2$ is the coefficient quantifying the degree of asymmetry (in the CM frame).
For the \kap\ efficiency estimation $a_2$ was set to 0.3, and 
$T_{\text{eff}}$ to $100\,\text{MeV}$. The observed differences were found to be
negligible ($< 2 \%$) 
and had no influence on the systematic uncertainty of the final results. 

\begin{figure*}
 \centering
\begin{tabular}{cc}
 \begin{minipage}[c]{.5\textwidth}
 \includegraphics[width=7.5cm]{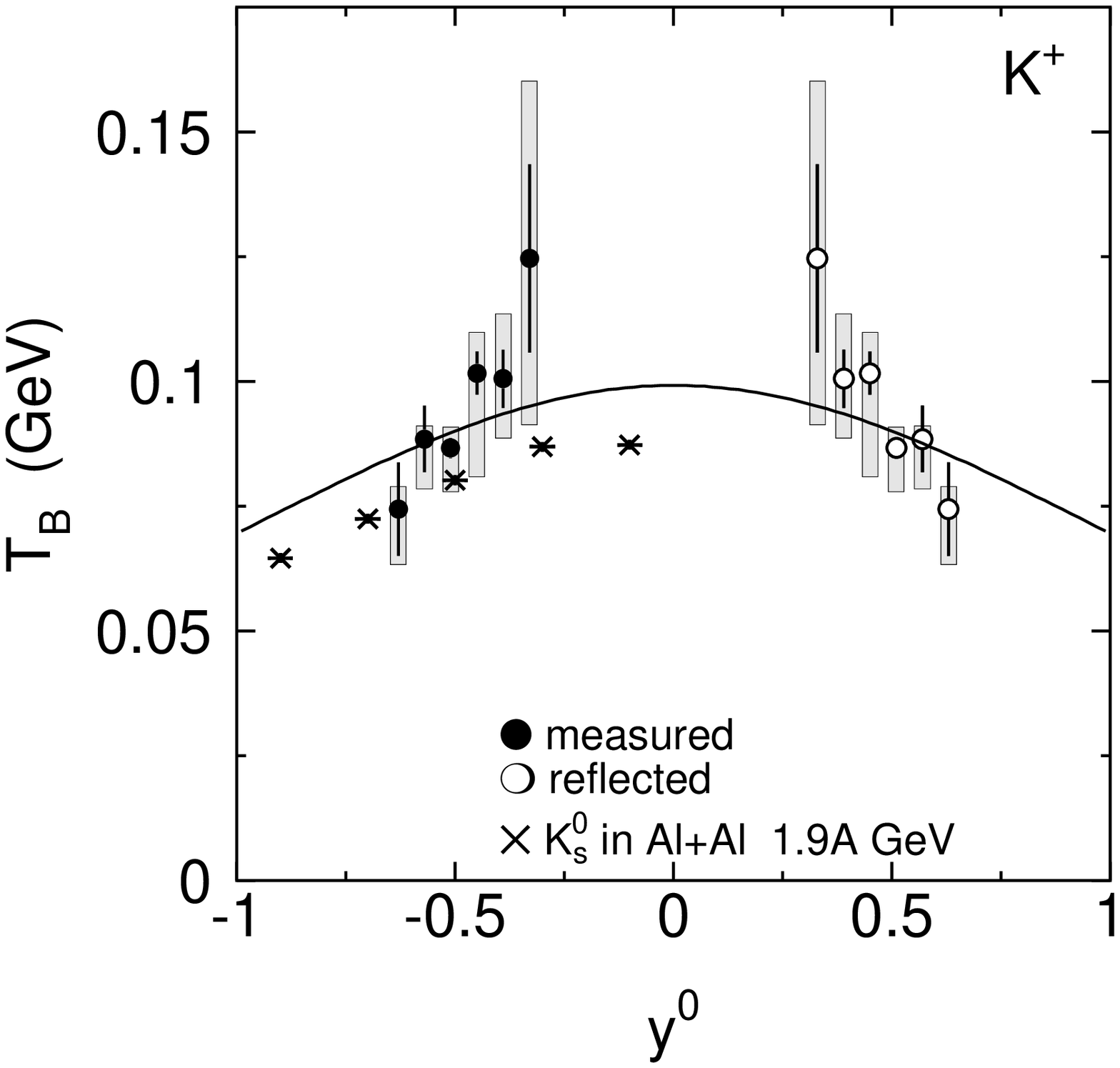}
 \end{minipage}
&
 \begin{minipage}[c]{.5\textwidth}
 \includegraphics[width=7.5cm]{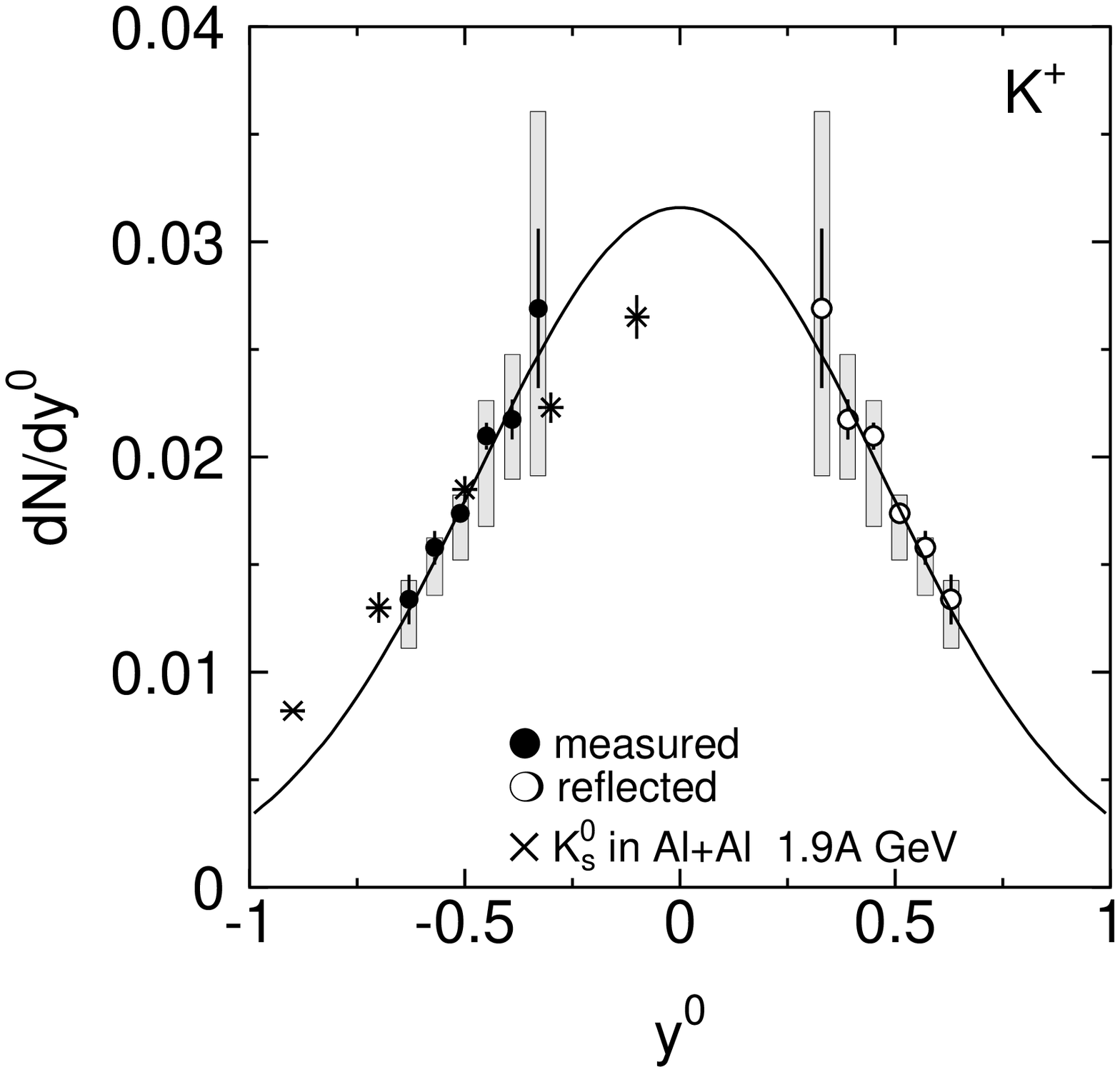}
 \end{minipage}
 \end{tabular}
 \caption{\label{fig:kptb} Left: inverse slope parameters ($T_\text{B}$) 
  of the transverse mass spectra of \kap\ mesons as a function of the 
  normalised rapidity. Solid line represents the fit of
  function~(\ref{eq:cosh}).  
  Right: rapidity density distribution for \kap. The solid
  curve represents the 
  Gaussian fit to the data. 
  Crosses depict the respective emission patterns of K$^0_\text{s}$
  from the same experiment~\cite{Lope10}.
  The experimental results (full circles) were reflected (hollow circles) 
  due to the symmetry of the colliding system. The vertical bars and
  shaded rectangles 
  correspond to the statistical and systematic uncertainties, respectively.}
\end{figure*}

\subsubsection{Efficiency of $\phi$ reconstruction}
\label{sec:fieff}

For the evaluation of the reconstruction efficiency $\phi$ mesons were
generated according to the isotropic ($a_2=0$) Boltzmann-like function
(c.f. eq.~(\ref{eq:EkThe})), and added to the IQMD events 
($\sim6\times10^5$ events were simulated in total). 
Since the precise kinematic properties of $\phi$ meson are unknown,  
the $T_\text{eff}$ parameter was varied within \mbox{[80, 130]}~MeV. 
The influence of these variations on the final results 
was included in the respective systematic errors. 
As all the simulations were performed in the 
$\phi \rightarrow \text{K}^+ \text{K}^-$ channel, the efficiency
values presented below will be given as normalised with respect to this channel.
However, for the final results reported in sec.~\ref{sec:fiyield}, 
the branching ratio of this decay mode will be included. The efficiency
analysis presented below was performed for the "standard" set of track
quality cuts, however, both "narrow" and "wide" sets were used to evaluate the
systematic uncertainties (see sec.~\ref{sec:kaonid} for more details). 

For more thorough understanding of the $\phi$ meson detection losses, 
the total efficiency ($\epsilon_\text{tot}$) was split into two terms
(see eq.~6 in~\cite{Mang03}): 

\begin{equation}
\label{eq:eff}
\epsilon_\text{tot} = \epsilon_\text{max} \cdot \epsilon_\text{det}\quad .
\end{equation}
$\epsilon_\text{max}$ is the efficiency of detection of
$\phi$ mesons in the K$^+$K$^-$ channel by the ideal apparatus
and includes losses due to the geometrical acceptance,
decays of kaons on their path towards ToF Barrel,
and the limitation of momenta of kaons to 0.6 GeV/$c$.
$\epsilon_\text{det}$ corresponds to the subsequent loss of $\phi$
meson signal due to the internal efficiency of the CDC+ToF Barrel
subsystem, including the limitations due to measurement capabilities,
tracking, matching quality, and the PID algorithm.

The same momentum cuts were applied in the efficiency evaluation procedure
as for the experimental data.
Table~\ref{tab:fieff} reports both the factors and the total efficiency 
for different temperatures of the $\phi$ meson sources. 
As expected, the $\epsilon_\text{det}$ was found to be independent 
within the 1$\sigma$ uncertainty of properties of the distribution
of primary $\phi$ mesons. 
Regarding $\epsilon_\text{max}$, in the analysis presented in~\cite{Mang03}
the variations of this parameter due to different temperatures of the 
$\phi$ meson source reached a factor of 4. In contrast, for our analysis
$\epsilon_\text{max}$ was found to be clearly better anchored. 
This is due to the different acceptance windows available for both experiments.
For the Al+Al case shifting the target by 40~cm upstream with respect 
to its nominal position resulted in moving the phase space coverage 
towards midrapidity, where the production yield is much greater.
The total $\phi$ meson efficiency was found to be about 0.4 -- 0.5~\%.

\begin{table}
\caption{Efficiencies for  $\phi$ meson detection in the 
   K$^+$K$^-$ decay channel for different temperatures of the $\phi$ emitting
   source.  
   $\epsilon_\text{max}$ is the efficiency within the geometrical 
   and momentum limits of an ideal apparatus including the kaon decays on 
   their path toward ToF Barrel, and $\epsilon_\text{det}$ is the efficiency 
   of the CDC+Barrel detector for kaons reaching the ToF Barrel (see
   eq.~(\ref{eq:eff})).  
   $\epsilon_\text{tot}$ is the total detection efficiency.}
\label{tab:fieff}
\begin{tabular}{ccccccc}
\hline\noalign{\smallskip}
$T_\text{eff}$ [MeV] & $\epsilon_\text{max}$ [\%] & $\epsilon_\text{det}$ [\%] & $\epsilon_\text{tot}$ [\%] \\
\noalign{\smallskip}\hline\noalign{\smallskip}
  80  &  0.88 $\pm$ 0.02  &  44 $\pm$ 2  &  0.38 $\pm$ 0.02 \\
 100  &  0.99 $\pm$ 0.02  &  43 $\pm$ 2  &  0.43 $\pm$ 0.02 \\
 120  &  1.03 $\pm$ 0.02  &  47 $\pm$ 2  &  0.48 $\pm$ 0.02 \\
 130  &  1.03 $\pm$ 0.02  &  45 $\pm$ 2  &  0.46 $\pm$ 0.02 \\
\noalign{\smallskip}\hline
\end{tabular}
\end{table}

\section{Results on the production of charged kaons}
\label{pga:kres}

\subsection{K$^+$ phase space analysis in the $m_\text{t} - y^0$ representation}
\label{sec:kpmt}
This section reports on the two-dimensional $m_\text{t} - y^0$ analysis of the
\kap\ phase space distribution, and the calculation of the total emission
yield. The same analysis applied to \kam\ was found to be unstable  
due to considerably lower statistics. 

\begin{figure*}
\begin{tabular}{cc}
 \begin{minipage}[c]{.5\textwidth}
 \includegraphics[width=7.5cm]{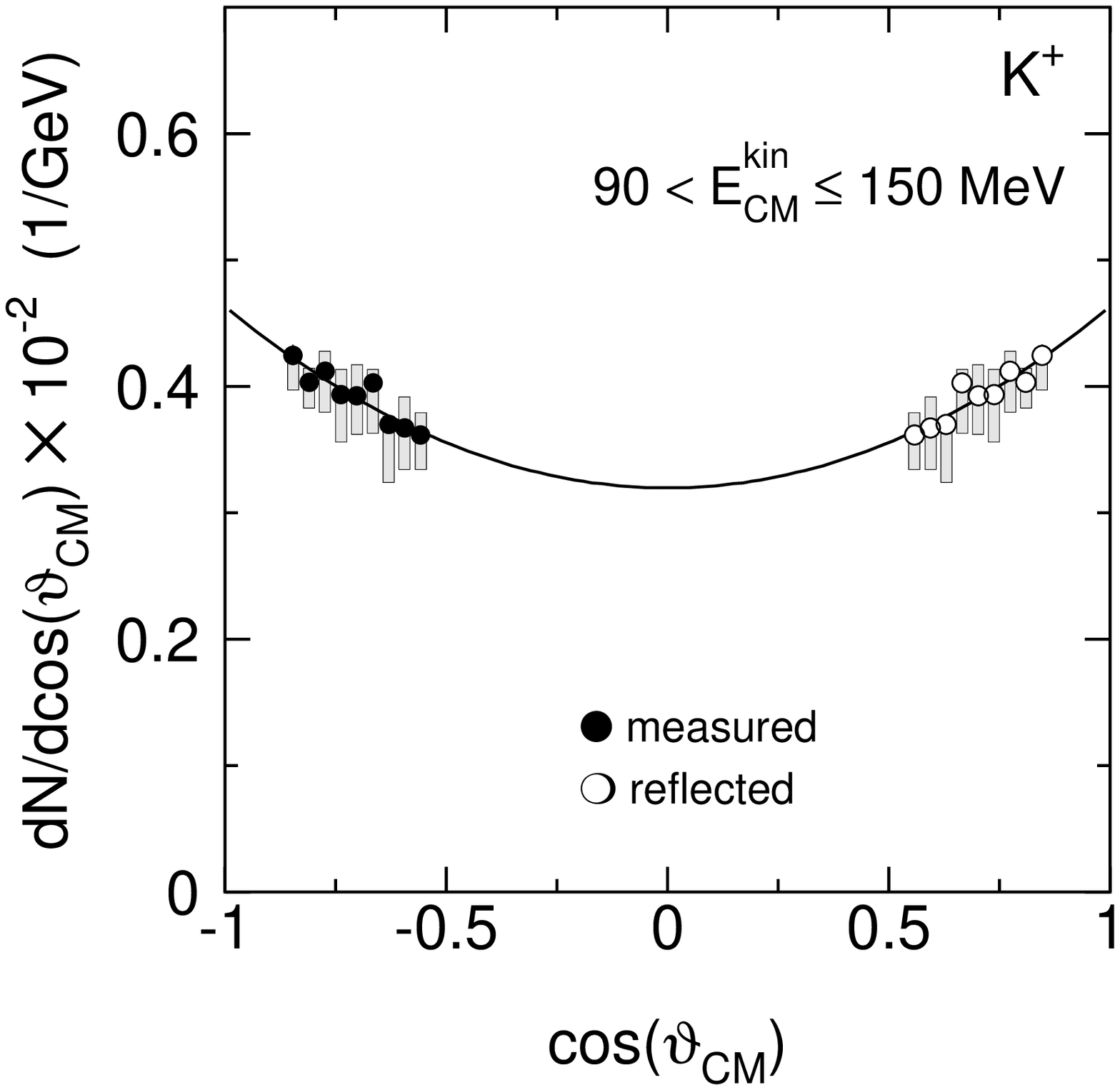}
 \end{minipage}
&
 \begin{minipage}[c]{.5\textwidth}
 \includegraphics[width=7.5cm]{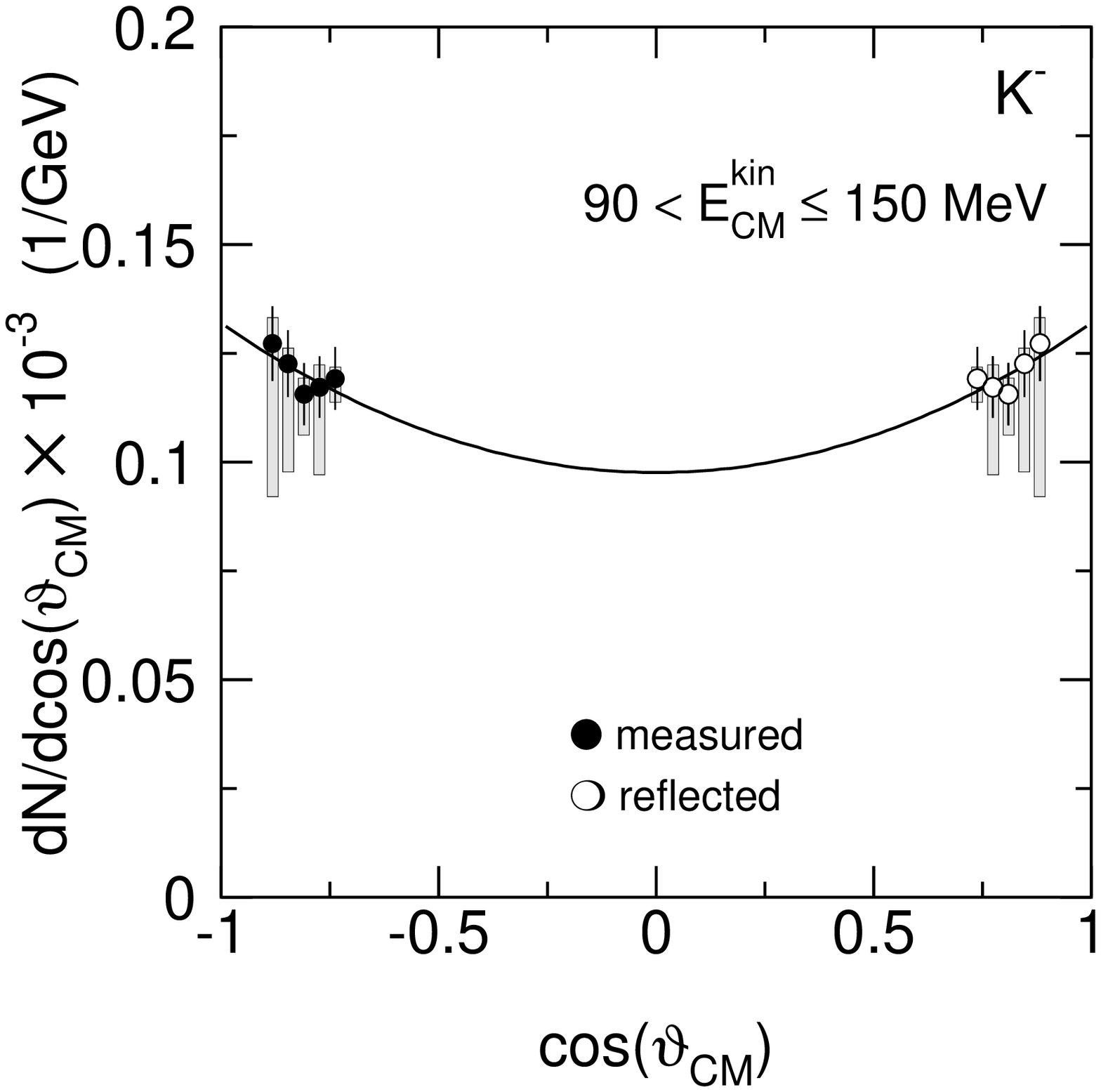}
 \end{minipage}
 \end{tabular}
 \caption{Angular distribution of positive (left) and negative (right) kaons
   plotted as a function of $\cos (\vartheta_\text{CM})$ for the particles with
   center-of-mass kinetic energy of $E_\text{CM}^\text{kin} \in [90, 150]$~MeV. Experimental values
   are reflected with respect to $\cos \vartheta_\text{CM} = 0$. The solid line
   represents the fit of function~(\ref{eq:EkThe_The}). Black bars and shaded
   rectangles correspond to the statistical and systematic uncertainties,
   respectively.}
 \label{fig:kpkmthe}
\end{figure*}

We start from the Boltzmann-like distribution pa\-ra\-me\-tri\-sed in 
$y_\text{CM}$ (rapidity of particle in the frame of its source),
and transverse mass $m_\text{t} = \sqrt{p^2_{\text{t}} + m_0^2}$ 
($m_0$ is the particle's mass in vacuum),

\begin{equation}
 \label{eq:mtymaster}
 \left. \frac{1}{m_\text{t}^2}\frac{\text{d}^2 N}{\text{d}m_\text{t}\text{d}y_\text{CM}} 
 \right|_{y_\text{CM,i}} 
  = A_0 (y_\text{CM}) \cdot e^{ -m_\text{t}/T_\text{B}(y_\text{CM}) } \quad ,
\end{equation}

\noindent where $A_0$ is the normalisation factor, and $T_{\text{B}}$ is the
inverse slope. In contrast to the pure Boltzmann approach (cf. Sect. III 
in~\cite{Schn93}), we allow $A_0$ and $T_{\text{B}}$ to be the free functions 
of rapidity, which are to be extracted by fitting to the experimental data. 
In the next step we change the variables to $m_\text{t} - m_0$, 
and $y^0$. For every $i$-th slice of $y^0$,

\begin{equation}
 \label{eq:mtfit}
 \left. \frac{1}{m_\text{t}^2}\frac{\text{d}^2 N}{\text{d}(m_\text{t} - m_0)\text{d}y^0} 
 \right|_{y^0_\text{i}} = A(y^0_\text{i}) \cdot e^{ -(m_\text{t} - m_0)/T_\text{B}
 (y^0_\text{i}) }  \quad ,
\end{equation}

\noindent where 
$A(y^0_\text{i}) = A_0(y^0_\text{i}) \cdot y_\text{NN} \cdot \exp (-m_0/T_\text{B})$. 
The reconstructed spectra of transverse mass of positively charged kaons within 
the $y^0$ range of $(-0.66, -0.36)$ were scaled by $1/m_\text{t}^2$, 
and plotted logarithmically in fig.~\ref{fig:kpmt}. 
The systematic errors are shown by the shaded rectangles. 
They reflect the sensitivity of data to 
variations of cut parameters in the procedures of kaon track finding, 
different parameters in the background evaluation procedure, and the 
uncertainty associated with the adjustment of the centrality class 
between the experimental and simulated data. The convention for these errors 
applied throughout the present paper is to cover all the observed range 
of the data point caused by variations of the above-mentioned parameters.
The spectra were then fitted with the eq.~\ref{eq:mtfit}, 
and $T_\text{B}$ parameters were found.
We do not imply the thermalisation of kaons, but use the above-mentioned 
formula as a suitable representation of the data 
and the $T_\text{B}$ as a measure of hardness of the spectrum. 

The rapidity distribution of the slope parameters $T_\text{B}$ extracted 
by the above-mentioned fitting procedure is \linebreak shown in the 
left panel of fig.~\ref{fig:kptb}.
The symmetry of the colliding system allows to reflect the data points
with respect to midrapidity.
The figure also shows the $T_\text{B}$ profile of K$^0_\text{s}$ 
emitted from the same colliding system~\cite{Lope10}. 
Spectra of positively charged kaons appear to be somewhat harder than those 
of K$^0_\text{s}$, however, large systematic errors do not exclude 
that the slopes are identical. 
The distribution of temperatures as a function of normalised rapidity has 
been fitted with the prediction of the Boltzmann  
model of particles emitted isotropically from the thermalised source:

\begin{equation}
 \label{eq:cosh}
 T_\text{B} (y^0) = \frac{T_\text{eff}}{\cosh ~(y^0 \cdot y_\text{NN}) } \quad ,
\end{equation}

\noindent where the temperature $T_\text{eff}$ within this model,
considered as an overall effective parameter quantifying the hardness 
of the kaon emission, has been found to be 
$T_\text{B} (\text{K}^+) = 
99 \pm 2 \text{(stat)} ^{+~4}_{-11} \text{(syst)} $~MeV.

\begin{figure*}
 \begin{tabular}{cc}
 \begin{minipage}[c]{0.5\textwidth}
 \includegraphics[width=7.5cm]{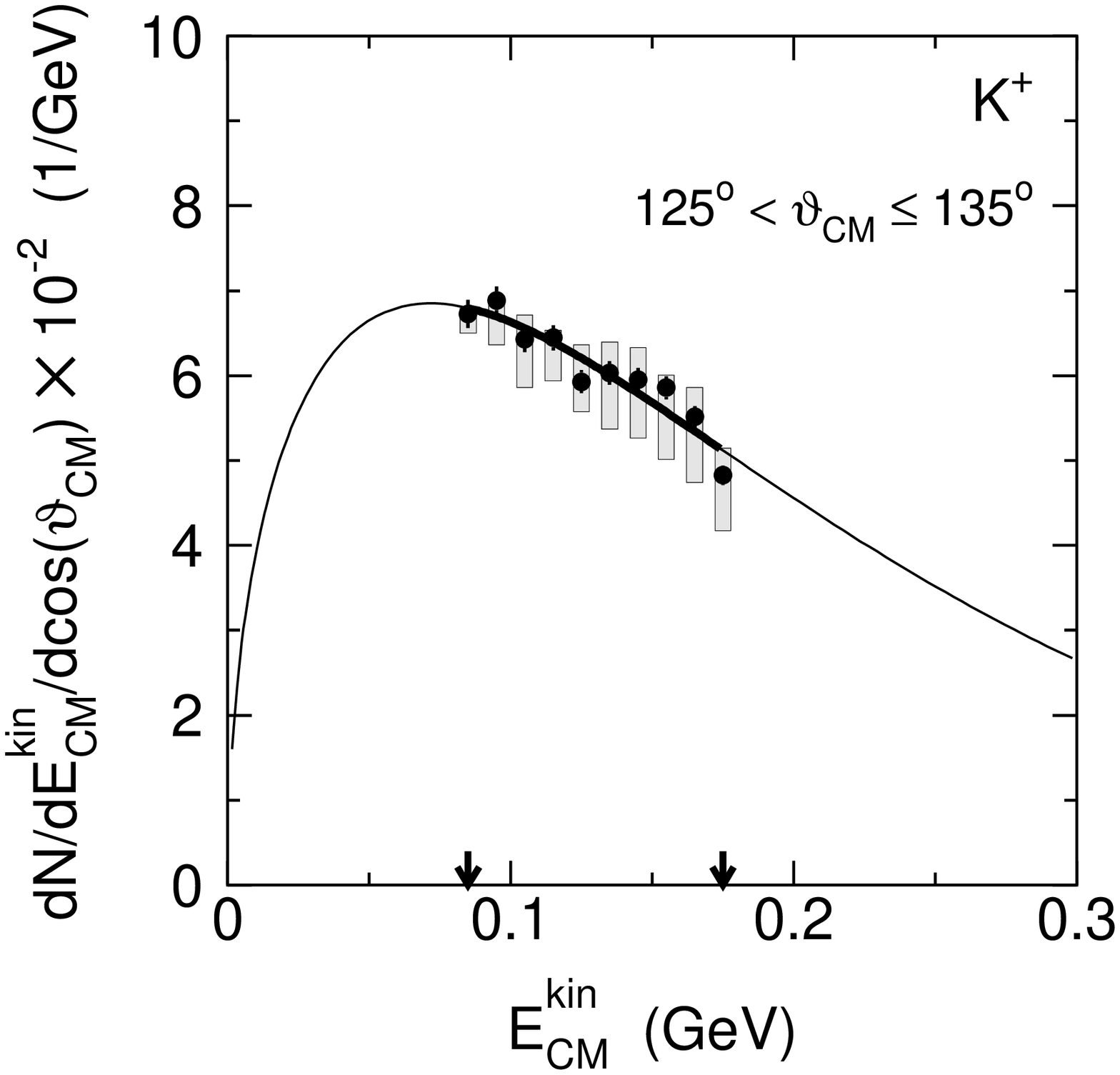}
 \end{minipage}
&
 \begin{minipage}[c]{0.5\textwidth}
 \includegraphics[width=7.5cm]{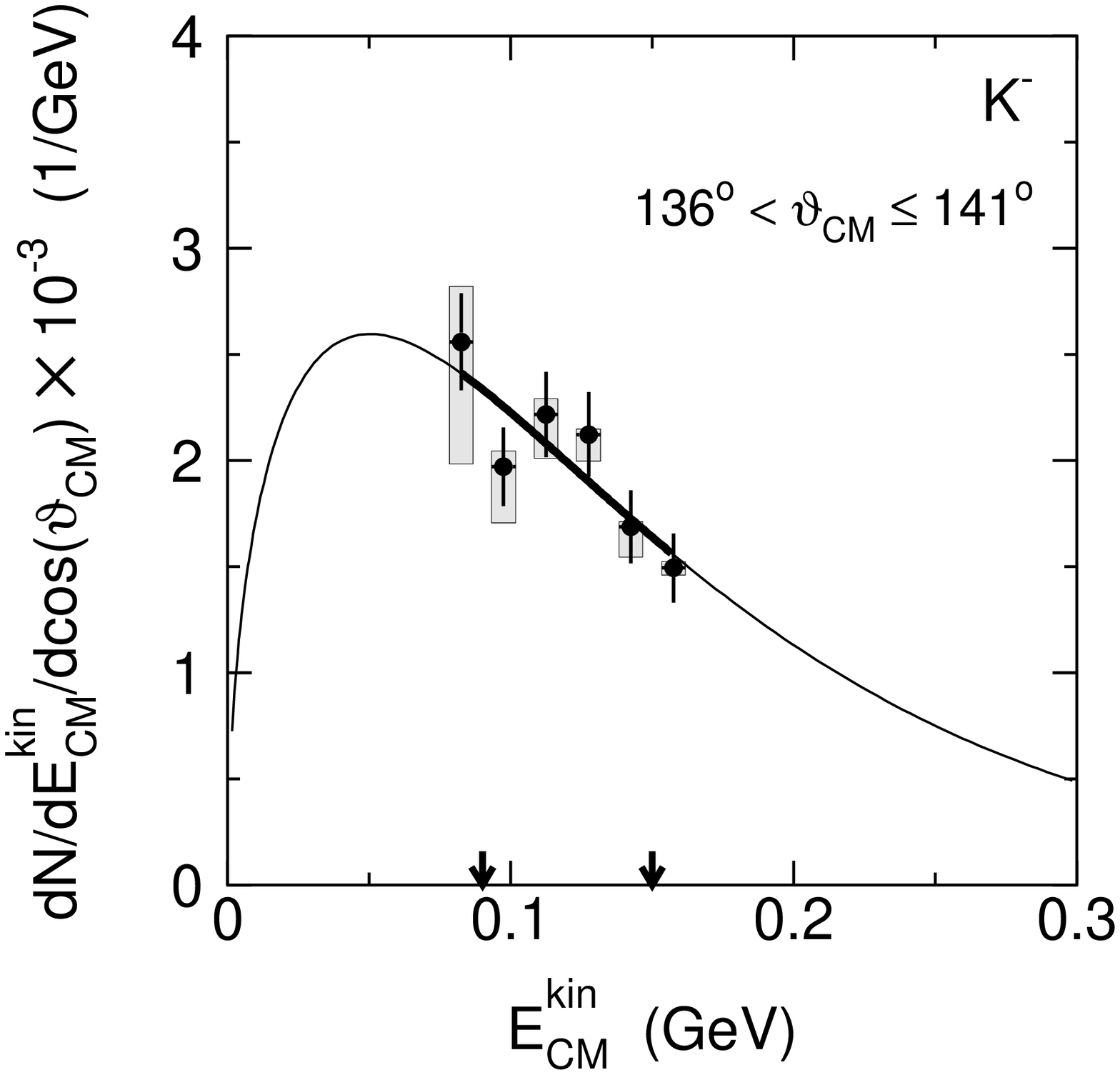}
 \end{minipage}
 \end{tabular}
 \caption{Kinetic energy distribution of charged kaons in the 
  CM frame obtained for the particles emitted within 
  $\vartheta_\text{CM} \in [125^\circ, 135^\circ]$ (K$^+$) and 
  $\vartheta_\text{CM} \in [136^\circ, 141^\circ]$ (K$^-$).
  The full line represents the fit of function~(\ref{eq:EkThe_Ek}). 
  The black arrows indicate the window used for adjustment of 
  normalisation to eq.~(\ref{eq:EkThe_The}), see text for details.} 
 \label{fig:kpkmekin}
\end{figure*}

The distribution of normalised rapidity of \kap\ has
been obtained by an analytic integration of eq.~(\ref{eq:mtfit})
from 0 to $\infty$, 

\begin{equation}
 \label{eq:dNdy0}
 \left. \frac{\text{d}N}{\text{d}y^0} \right|_{y^0_\text{i}} 
 = A \cdot T_\text{B}^3 \cdot \left( \frac{m_0^2}{T_\text{B}^2} 
 + 2\frac{m_0}{T_\text{B}} + 2\right) \quad .
\end{equation}

\noindent In the evaluation of the statistical error of this formula, 
the covariance term between $A$ and $T_\text{B}$ parameters was included.
The resulting d$N$/d$y^0$ distribution is shown in fig.~\ref{fig:kptb}.
In the range of rapidities available for this analysis
the profiles for the K$^+$ and K$^0_\text{s}$ from ref.~\cite{Lope10} are
found to overlap. 

The analysis of the two-dimensional $m_\text{t} - y^0$
distribution allows to extrapolate the data from an experimentally
available region of phase space, and integrate the
distribution to obtain the total yield. However, since the
accessible region is partial, the extrapolated value of yield
may depend on the model assumptions. The data obtained within the ToF
Barrel acceptance window follow a Gaussian profile which is shown 
by the full line in the right panel of fig.~\ref{fig:kptb}. 
We stress here, that this profile is employed only
phenomenologically. A cross-check of this approach by
reconstructing the yield using different representation
will be presented in the next subsection.

The total K$^+$ emission yield within this approach
is \linebreak$P_{\text{K}^+} = 
 \left[ 3.72 \pm 0.05 (\text{stat}) ^{+0.28}_{-0.52} (\text{syst}) \right] 
 \times 10^{-2}$. 
\noindent Interestingly, the obtained yield is identical within experimental 
uncertainties to that of K$^0$, extracted using the same $m_\text{t}-y^0$ 
analysis, although in the considerably wider range of phase space~\cite{Lope10}.

\subsection{Kinetic energy and polar angle distributions of \kap\ and \kam}
\label{sec:kpmekin}

In this approach 
it is assumed that the charged kaon emission in the nucleon-nucleon centre
of mass can be parametrised by the Boltzmann-like kinetic energy term multiplied 
by the polar-angular term (c.f. eq.~(\ref{eq:EkThe})). 
We only note that in formula~(\ref{eq:EkThe}) the kinetic energy and
polar angle are not correlated. This allows to extract the $a_2$ and $T_\text{eff}$ 
parameters independently by projecting data on the $\cos \vartheta_\text{CM}$ and 
$E_\text{CM}^\text{kin}$ axes, correspondingly. In light of the limited sample 
of negative kaons, the fact that the total number of projections in this 
method is smaller than for the $m_\text{t} - y^0$ analysis turned out to be 
crucial for stabilising the \kam\ results. 

The spectra of kinetic energy and polar angle were trimmed to the available
phase space (c.f. figs.~\ref{fig:kp}  
and~\ref{fig:km}) in order to minimise the edge effects.

Fig.~\ref{fig:kpkmthe} shows the polar angle distributions of \kap, and \kam\
emitted with kinetic energies $90 < E_\text{CM}^\text{kin} < 150$~MeV, and in the 
angular range: $123^\circ < \vartheta_\text{CM} < 150^\circ$ 
($136^\circ < \vartheta_\text{CM} < 154^\circ$) for \kap\ (\kam), respectively.
The corresponding projection of eq.~(\ref{eq:EkThe}) runs as follows:

\begin{equation}
  \label{eq:EkThe_The}
  \left[ \frac{\text{d}N}{\text{dcos}\vartheta_\text{CM}} \right] _{E_\text{k1} .. E_\text{k2}} = 
       ~B_{E_\text{k1} .. E_\text{k2}} \cdot (1 + a_2 \cos^2 \vartheta_\text{CM})\quad ,
\end{equation}

\noindent where $B_{E_\text{k1} .. E_\text{k2}}$ is the normalisation 
constant for kaons with kinetic energies in the range $[E_\text{k1} .. E_\text{k2}]$
(here, \linebreak$[90 .. 150]$~MeV). 
A fit of this formula to the angular distributions of \kap\ and \kam\ mesons
allowed to extract the respective $a_2$ coefficients (see
Table~\ref{tab:kpkmresults} for all the relevant results of this procedure):  

\begin{equation}
 \label{eq:angresult}
 \begin{matrix}
 a_2 (\text{K}^+) & = 0.45 \pm 0.08 (\text{stat}) ^{+0.18}_{-0.11} (\text{syst}) &       \\ 
 a_2 (\text{K}^-) & = 0.35 \pm 0.40 (\text{stat}) ^{+0.11}_{-0.83} (\text{syst}) &\quad .\\
 \end{matrix}
\end{equation}

\noindent 
The obtained coefficients were found to be rather small for \kap, and
consistent with 0 for \kam, a pattern observed also in the somewhat heavier Ni+Ni 
system at the same beam energy (see fig.~16b in~\cite{Fors07}).
Values of $B_{E_\text{k1} .. E_\text{k2}}$ extracted by the fit procedure 
are reported in table~\ref{tab:kpkmresults}, and
will be further used for the evaluation of the total yield. 

\begin{table*}
\caption{Parameters of the \kap\ and \kam\ phase space distributions
obtained by fitting the $E_\text{CM}^\text{kin}$ and $\vartheta_\text{CM}$ projections of 
eq.~(\ref{eq:EkThe}): $a_2$ (polar angle anisotropy parameter), 
$B_{E_\text{k1} .. E_\text{k2}}$ (normalisation parameter in 
eq.~(\ref{eq:EkThe_The})), $I_{E_\text{k1} .. E_\text{k2}}$ 
(partial integral, see eq.~(\ref{eq:EkThe_TheInt})), $T_\text{eff}$ 
(effective temperature), and $P$ (total production yield). The first 
and second error denote the statistical and systematic errors, respectively.
}
\label{tab:kpkmresults}
\begin{tabular}{cccccc}
\hline\noalign{\smallskip}
      & 
$a_2$                                & 
$B_{E_\text{k1} .. E_\text{k2}}$ &
$I_{E_\text{k1} .. E_\text{k2}}$ &  
$T_\text{eff} ~\text{[MeV]}$       & 
P                                      \\
\noalign{\smallskip}\hline\noalign{\smallskip}
K$^+$ &
$0.45 \pm 0.08 ^{+0.18}_{-0.11}$                  &
$[3.2 \pm 0.1 ^{+0.2}_{-0.5}] \times 10^{-3}$     &
$[7.35 \pm 0.08 ^{+0.33}_{-0.73}] \times 10^{-3}$ &
$109 \pm 2 ^{+6}_{-13}$                           &
$[3.75 \pm 0.07 ^{+0.33}_{-0.64}] \times 10^{-2}$   \\

K$^-$ &
$0.35 \pm 0.40 ^{+0.11}_{-0.83}$                  &
$[9.8 \pm 2.1 ^{+5.0}_{-0.3}] \times 10^{-5}$     &
$[2.18 \pm 0.21 ^{+0.30}_{-0.03}] \times 10^{-4}$ &
$~~82 \pm 6 ^{+21}_{-6}$                          &
$[9.5 \pm 1.0 ^{+2.6}_{-0.1}] \times 10^{-4}$   \\
\noalign{\smallskip}\hline
\end{tabular}
\end{table*}

In the next step, the kaon data in the polar angle ranges of
$125^\circ < \vartheta_\text{CM} < 135^\circ$ (for \kap), and
$136^\circ < \vartheta_\text{CM} < 141^\circ$ (for \kam) were projected 
to the kinetic energy axis. The resulting $E_\text{CM}^\text{kin}$ distributions 
are shown in fig.~\ref{fig:kpkmekin}. 
The corresponding projection of eq.~(\ref{eq:EkThe}) follows:

\begin{equation}
  \label{eq:EkThe_Ek}
  \begin{split}
  \left[ \frac{dN}{dE_\text{CM}^\text{kin}} \right] 
        &_{\vartheta_{\mathrm{CM},1} .. \vartheta_{\mathrm{CM},2}} =  \\
        & ~C_{\vartheta_{\mathrm{CM},1} .. \vartheta_{\mathrm{CM},2}} 
          \cdot p_\mathrm{CM}E_\mathrm{CM} \exp (-E_\mathrm{CM} /T_\text{eff})\quad ,
  \end{split}
\end{equation}

\noindent where $C_{\vartheta_{\mathrm{CM},1} .. \vartheta_{\mathrm{CM},2}}$ is the normalisation constant 
for kaons emitted in the polar angle range of $[\vartheta_{\mathrm{CM},1} .. \vartheta_{\mathrm{CM},2}]$. 
Fitting the formula above to the experimental data allowed to extract the
following inverse slopes:

\begin{equation}
\label{eq:teff_kpkm}
 \begin{matrix}
 \quad\quad T_\text{eff} (\text{K}^+) & = 109 \pm 2 (\text{stat}) ~^{+~6}_{-13} (\text{syst}) \,\text{MeV} &\\ 
 \quad\quad T_\text{eff} (\text{K}^-) & =~~82 \pm 6 (\text{stat}) ~^{+21}_{-~6} (\text{syst}) \,\text{MeV} &\quad . \\
 \end{matrix}
\end{equation}

These results are in a good agreement with the inverse slopes of charged kaons
emitted from Ni+Ni collisions in a similar centrality class
($\langle A_\text{part} \rangle_\text{b} = 34$),
both in terms of absolute values and in the superiority of $T_{\text{K}^+}$ over
$T_{\text{K}^-}$ (c.f. figs.~15b and 20 in~\cite{Fors07}). 

\subsection{Total emission yields of \kap\ and \kam}

In order to obtain the total kaon emission yield 
from the results reported in sec.~\ref{sec:kpmekin}, 
one has to account for different ranges of scope 
in the eqs.~(\ref{eq:EkThe_The}) and (\ref{eq:EkThe_Ek}). 
Thus, in the first step the integral of eq.~(\ref{eq:EkThe_The}) was evaluated:

\begin{equation}
 \label{eq:EkThe_TheInt}
 I_{E_\text{k1} .. E_\text{k2}} =
 \int_{-1}^{1}
 \left[ \frac{\text{d}N}{\text{dcos}\vartheta_\text{CM}} \right] _{E_\text{k1} .. E_\text{k2}} 
 = 2B_{E_\text{k1} .. E_\text{k2}} \cdot \left(1+\frac{a_2}{3}\right)\quad .
\end{equation}

\noindent The results are presented in table~\ref{tab:kpkmresults}. 
This integral represents the yield of kaons emitted in the $E_\text{CM}^\text{kin}$
range of \mbox{[90, 150]}~MeV. 
Note, that the smaller relative errors of $I_{E_\text{k1} .. E_\text{k2}}$ 
compared to those for $B_{E_\text{k1} .. E_\text{k2}}$ and $a_2$ 
fit parameters are due to strong anti-correlation term between the latter parameters. 
In the next step the normalisation of the kinetic energy spectrum expressed by
eq.~(\ref{eq:EkThe_Ek}) was adjusted such that the integral of this spectrum
in range $E_\text{CM}^\text{kin} \in [90, 150]~\text{MeV}$ was equal to the value of
$I_{E_\text{k1} .. E_\text{k2}}$. 
Such an adjusted energy spectrum was subsequently integrated to obtain 
the total yield of charged kaons:

\begin{equation}
 \label{eq:yield_kpkm}
 \begin{matrix}
\quad &P_{\text{K}^+} = & \left[3.75  \pm 0.07 (\text{stat}) ^{+0.33}_{-0.64} (\text{syst})  \right] \times 10^{-2} &\\[2ex]
\quad &P_{\text{K}^-} = & \left[0.95  \pm 0.10 (\text{stat}) ^{+0.26}_{-0.01} (\text{syst})  \right] \times 10^{-3} &\quad .\\ 
 \end{matrix}
\end{equation}

\noindent As mentioned above, this method provided a stable value of the \kam\ yield. 
The yields of \kap\ obtained using this approach, and by the 2-dimensional 
analysis of the $m_\text{t}-y^0$ phase space were found to be consistent. 
The systematic errors, as for all the other results mentioned above, 
account for the variations of cutting parameters in the procedures of 
kaon track finding and background subtraction, 
and the determination of the collision centrality. They cover all the range of the yield values due to these variations.

In comparison to the charged kaon multiplicities at similar energy and centrality,
the obtained yields of \kap\ (\kam) were found
to be about 1.8 (2.2) times smaller than for kaons emitted in Ni+Ni collisions 
at $\langle A_\text{part} \rangle_\text{b} = 74$~\cite{Menz00}. 
They were also found to be about 2.1 (1.4) times larger than for kaons emitted from
Ni+Ni at the kinetic energy of 1.8A GeV, and for a similar centrality class,
parametrised by $\langle A_\text{part} \rangle_\text{b} = 37.5$~\cite{Bart97}.

\begin{figure*}[tbp]
 \centering
\includegraphics[width=12cm]{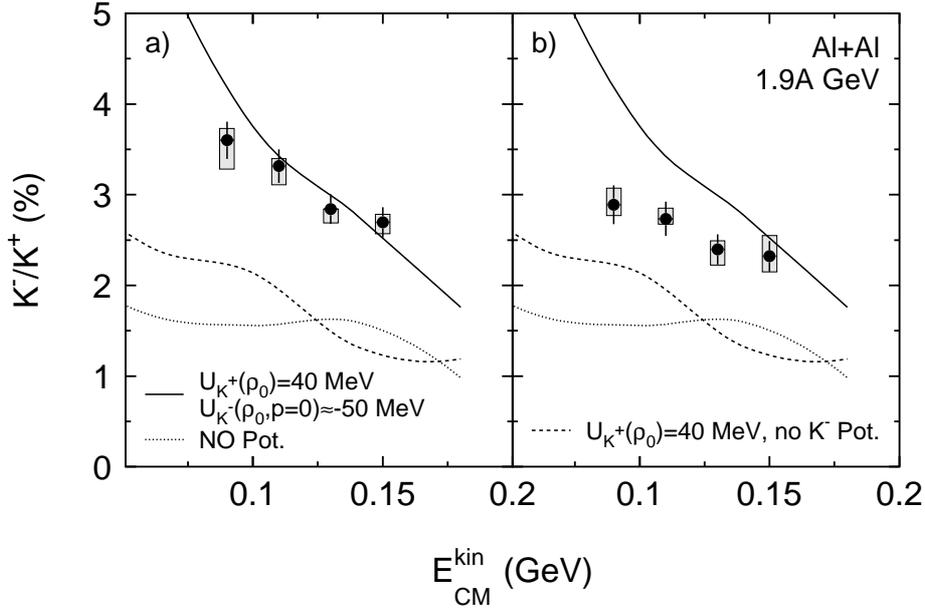}\\
 \caption[]{a) the K$^-$/K$^+$ ratio as a function of $E_\text{CM}^\text{kin}$ in the Al+Al
  experiment. Error bars represent the statistical uncertainties. Shaded rectangles
  represent the estimation of systematic errors. Lines represent the
  results of HSD model predictions including different values of KN potentials. 
  b) the K$^-$/K$^+$ ratio distribution corrected for \kam\ mesons from $\phi$ decays.
See sec.~\ref{sec:kmfromphi} for more details.}
 \label{fig:kmkpratio}
\end{figure*}

\subsection{\kam/\kap\ ratio as function of kinetic energy}
\label{sec:kaonhsd}
The \kam/\kap\ ratio has been studied as a function of the kinetic energy of 
kaons and their rapidity. 
Studying particle yield ratio offers two advantages \cite{Wisn00}. 
(i)  Experimental biases like detection efficiencies and acceptance 
     losses, are minimised. 
(ii) In-medium modifications of kaonic properties are predicted to generate
     the opposite kinematic effects on \kam\ and \kap, hence the ratio should 
     reveal them more clearly (see sec.~\ref{sec:disc} for 
     detailed discussion). 

Fig.~\ref{fig:kmkpratio}a shows the measured \kam/\kap\ ratio as a 
function of the kinetic energy in the c.m. reference frame 
($E_\text{CM}^\text{kin}$) for the Al+Al collisions. 
The polar-angle range 136$^\circ < \vartheta_\text{CM} < $150$^\circ$ has been 
chosen in order to reduce the edge effects of the CDC+Barrel subsystem 
and provide the largest possible scope of $E_\text{CM}^\text{kin}$. 
The black points indicate the experimental data. 
The error bars represent the statistical uncertainties, and the shaded 
rectangles depict the range of systematic errors. The ratio has been found to
decrease with energy $E_\text{CM}^\text{kin}$,  
which is a direct consequence of the lower transverse slope 
for \kam\ compared to that of \kap. A comparison of this distribution
with HSD transport model predictions is discussed in sec.~\ref{sec:disc}.

\begin{figure*}[tbp]
 \centering
 \includegraphics[width=7.5cm]{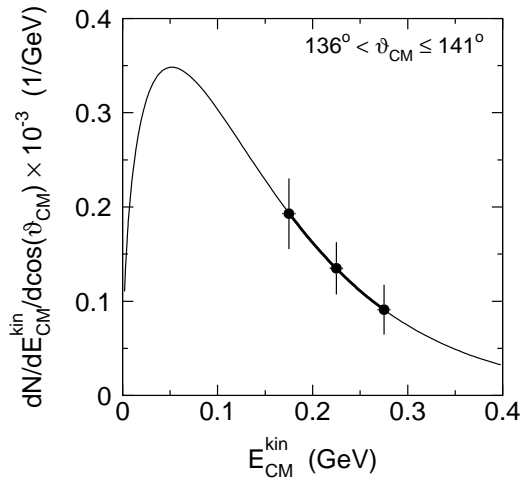}
 \caption[]{Kinetic energy distribution of $\phi$ mesons emitted in polar
   angles $\vartheta_\text{CM} \in \left[ 115^{\circ} , 145^{\circ} \right]$. 
   The full line depicts the Boltzmann function fit.

   }
 \label{fig:fiekin}
\end{figure*}

\section{Results on $\phi$ mesons}
\label{sec:fiyield}

A sample of about 108 $\phi$ mesons is too small for a detailed analysis 
of phase space of this particle. In order to obtain the emission yield in 
the \kap\kam\ channel the events were summed up over the available phase 
space region, and divided by the detection efficiency reported in 
sec.~\ref{sec:fieff}. The total yield per triggered event was obtained 
by dividing the result above by the branching ratio of the 
$\phi \rightarrow \text{K}^+ \text{K}^-$ channel, and was found to be: 

\begin{equation}
 \rm{P(\phi) = [3.3 ~\pm~ 0.5 ~\text{(stat)} ~^{+0.4}_{-0.8} ~\text{(syst)}] 
     \times 10^{-4}}\quad .
 \label{eq:yield_phi}
\end{equation}

\noindent The systematic errors include different selection criteria
for good tracks, CDC+Barrel matching criteria, and widths of ellipse mass cuts.
They also account for the variations of parameters of the $\phi$ meson 
emission source assumed in the efficiency simulations, of the invariant 
mass range used for the normalisation of background, and cuts on charged 
particle multiplicities used for the collision centrality determination. 
As before, the systematic errors aim to cover all the observed range of the 
yield values due to variations of the above-mentioned parameters.
The result seems to be slightly higher than the yield of 
$(2.6 \pm 0.7) \times 10^{-4}$ obtained for the Ar+KCl collisions
(although within 0.7 standard deviation they could be equal), 
at a very similar $\langle A_\text{part} \rangle_\text{b}$, 
but at lower incident energy 1.756A~GeV~\cite{Agak09}. 
The obtained yield also seems to be lower than $(4.4 \pm 0.7) \times 10^{-4}$ 
measured for Ni+Ni collisions at the same beam kinetic energy of 1.91A~GeV, 
but at slightly higher $\langle A_\text{part} \rangle_\text{b}$
(although within 1.5 standard deviation both results could be equal) 
~\cite{Pias15}.

The kinetic energy distribution was also investigated. To minimise 
the side effects at the edges of the ToF Barrel detector, the acceptance window 
was trimmed by requiring $115^\circ < \vartheta_\text{CM} < 145^\circ$. 
Fig.~\ref{fig:fiekin} shows the measured spectrum. 
Eq.~(\ref{eq:EkThe_Ek}) was fitted to this spectrum, 
and the inverse slope was found to be:

\begin{equation}
 T_\text{eff} = 93 \pm 14~ \text{(stat)} ^{+17}_{-15} \text{(syst)}~\text{MeV}\quad ,
 \label{eq:fitemp}
\end{equation}

\noindent where the systematic errors include the same contributions
as for the total $\phi$ meson yield. 
The obtained slope seems to be slightly higher than that for the $\phi$ 
mesons from the Ar+KCl collisions, $84 \pm 8$~MeV, 
and slightly lower than the slope for the Ni+Ni collisions, 
$106 \pm 18$~MeV (although in both cases the standard deviation is 0.6).

\subsection{Influence of $\phi$ meson decays on the total \kam\ yield}
\label{sec:fitokm}

The yields of $\phi$ and \kam\ mesons for this colliding system
are of the same order of magnitude, and their ratio was deduced to:

\begin{equation}
 \label{eq:fi2km}
 \rm{\frac{P(\phi)}{P(\text{K}^-)} = 0.34 \pm 0.06 (stat) ^{+0.04}_{-0.14} (syst) 
     \quad .} 
\end{equation}

\noindent The systematic errors were calculated as in the previous
subsection, additionally taking into account that each of tested cases of event and
track selections was  
applied simultaneously in the evaluation procedures of the $\phi$ and \kam\ yields.
This result is found to be similar to the case of the Ar+KCl 
collisions ($37 \pm 13 \%$), and the Ni+Ni case
($44 \pm 7 \%$), however, different from the value of about unity 
observed for the elementary pp collisions 
at 2.65--2.85~GeV~\cite{Bale01,Maed08}).
Taking into account the $\phi \rightarrow $\kap\kam\ branching ratio
of 48.9\%, the result above means that $\left(17 \pm 3 ^{+2}_{-7}\right)\%$ of 
emitted \kam\ mesons originate from $\phi$ meson decays. 
To sum up, it appears that in the collisions of heavy-ions at beam energies 
of 1.7-1.9A~GeV the $\phi/ \text{K}^-$ ratio 
is about 40\%, and thus the contribution of $\phi$ decays into the
\kam\ emission yield of about 20\% is a general observation. 
The influence of the $\phi$ mesons on the kaon dynamics will be
discussed further in sec.~\ref{sec:kmfromphi}.

\section{Discussion}
\label{sec:disc}

Within the framework of transport models like IQMD or HSD~\cite{IQMD,HSD},
it is established that the initial kinematic distributions
of K$^\pm$ mesons are modified by the interplay between re-scattering,
absorption (mostly affecting \kam) and in-medium modification of kaonic
properties by the KN potential in dense medium. 

According to the IQMD and HSD calculations performed for heavy-ion 
collisions at 1-2A GeV, the first mechanism should increase the 
slopes of the kinetic energy distributions of kaons of both signs.
Absorption, by filtering out more \kam\ mesons with lower momenta, 
should effectively rise the slope, while the KN potential should
accelerate \kap, and decelerate \kam~\cite{Hart11}. 
The calculations within the transport models 
agree with the general experimental finding of superiority of
$T_{\text{K}^+}$ over $T_{\text{K}^-}$ inverse slopes, also observed 
for the Al+Al system (c.f. eq.~(\ref{eq:teff_kpkm})). 

Generally, it has been shown that transport models are able to describe the
kaon (and hyperon) yields and spectra when incorporating the in-medium 
modifications of kaon properties~\cite{Hart11}. Often the mass shifts of kaons 
in the medium are modelled inside transport codes by applying a parametrisation 
of the mass as a function of density. This approach is used in HSD for 
\kap\ mesons, while \kam\ mesons are treated as off-shell particles 
using the G-Matrix formalism \cite{Brat03}.

\subsection{\kam/\kap\ ratio compared to the HSD model}

It has been demonstrated that the ratio of kinetic energy distributions 
of \kam\ to \kap\ mesons is sensitive to KN potentials calculated within
the transport models~\cite{Wisn00,Fors03,RBUU1,RBUU2,Hart03}.
Fig.~\ref{fig:kmkpratio}a shows this ratio for the 
central Al+Al collisions. The ratio was investigated within the HSD model~\cite{HSD}. 
The profile obtained without any in-medium effects is shown 
as dotted line in fig.~\ref{fig:kmkpratio}, and falls 
far off the experimental data both in terms of slope and yield. 
If only K$^+$N potential with a linear dependence on density with 
$U_{\text{K}^+\text{N}}(\rho_0) = 40$~MeV is included 
in the calculations, the resulting profile falls slightly closer 
to the experimental data, as shown by the dashed line, 
but the agreement is still not reached. 
Finally, switching the in-medium effects for both kaons and antikaons
($U_{\text{K}^+\text{N}}(\rho_0) = 40$~MeV and 
($U_{\text{K}^-\text{N}}(\rho_0, p=0 ) \approx -50$~MeV)
results in the profile depicted by the solid line, reproducing
the experimental profile. 
Thus, a comparison of our data to the transport model predictions
leads qualitatively to the same conclusion as obtained in the central 
Ru+Ru/Zr collisions at 1.69A~GeV, and Ni+Ni collisions at 
1.93A~GeV~\cite{Wisn00}.

However, in the transport models mentioned above the influence of $\phi$
meson decays is missing (IQMD) or underestimated (HSD). 
In the following subsection we present an estimation of the range 
of this influence from the experimental Al+Al data.

\subsection{Influence of $\phi$ mesons on the kaon dynamics}
\label{sec:kmfromphi}

As shown in sec.~\ref{sec:fitokm}, the \kam\ emission has (at least)
two distinct contributions. While negative kaons emitted directly 
from the collision zone (dubbed below as "direct") reflect the conditions
therein, and may carry signatures of modifications of their properties 
in medium, most \kam\ mesons from $\phi$ meson decays are produced outside
the collision zone, at $\rho = 0$. 
As $m_{\phi}$ is only 32 MeV larger than $m_{\text{K}^+} + m_{\text{K}^-}$,
the pair of kaons originating from the $\phi$ meson decay receives only 
a small portion of energy from the energy balance, 
and \kam\ takes only half of this energy. It is therefore natural to expect
that the kinematics of these two sources of \kam\ emission is different.
On the other hand, as the yield of \kap\ is much higher to that of \kam\, 
the influence of $\phi$ meson decays on these particles is negligible. 

\begin{figure*}[tbp]
 \centering
 \includegraphics[height=7cm]{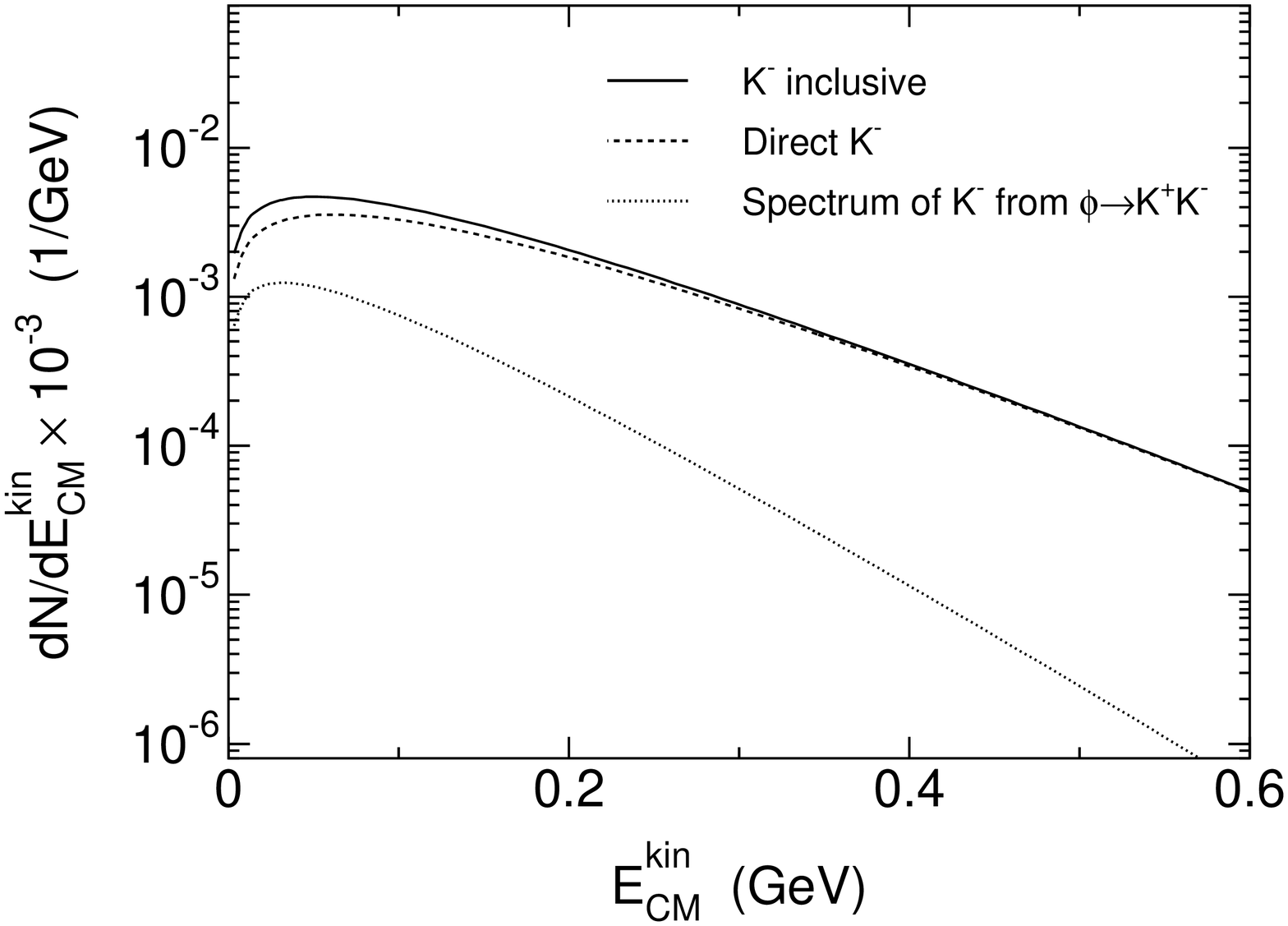}
 \caption{Kinetic energy distributions of \kam\ mesons obtained by the
   inclusive fit to the experimental data (solid line), \kam\ from $\phi$ 
   meson decays (dotted line), and "direct" \kam\ mesons generated from the
   thermal source (dashed line). See text for details.}
 \label{fig:kmfromphi}
\end{figure*}

In order to find the inverse slope of kinetic energy spectrum of negative
kaons emitted from $\phi$ meson decays, the latter were generated using the
PLUTO code~\cite{PLUTO}. Their emission was simulated according to the
Boltzmann distribution with temperature as reported in eq.~(\ref{eq:fitemp}),
where statistical and systematic uncertainties were taken into account. $\phi$
mesons were further permitted to decay into the \kap\kam\ pairs. The obtained
$E^\text{k}_\text{CM}$ distribution of \kam\ mesons was fitted with the
Boltzmann function and the inverse slope was found to be: 

\begin{equation}
 T_{\text{eff},\,\text{K}_\phi} = 
   55.5 \pm 7.1 \text{(stat)} ^{+8.5} _{-7.7} \text{(syst)} \,\text{MeV} \quad .
 \label{eq:teffkfi}
\end{equation}

The influence of binning of the energy
spectra of particles involved did not change the result significantly.

The inverse slope obtained in this way is clearly lower than that extracted from 
the inclusive fit to the \kam\ energy spectrum, c.f. eq.~(\ref{eq:teff_kpkm}).
As in the investigated colliding system about 17\% of produced \kam\ mesons
originate from the decays of $\phi$, one can put forward the hypothesis
that the inverse slope of the direct kaons should be greater than the 
inclusive value. It would imply that 
the feeding of negative kaons by $\phi$ meson decays is co-responsible
for enhancing the gap between the inverse slopes of \kam\ and \kap\ 
energy spectra (eq.~(\ref{eq:teff_kpkm})). 
A qualitative support for this reasoning was found in Ar+KCl collisions at
1.76A~GeV~\cite{Lore10}, and Ni+Ni collisions at 1.91A~GeV~\cite{Pias15}.
In order to verify this hypothesis, a double-source model of 
\kam\ emission from the investigated collision system was constructed. 
It was assumed that: 

\begin{itemize}
 \item the inclusive \kam\ emission spectrum is isotropic (an assumption
       justified by the value of the $a_2$ parameter comparable to 0 
       within 1 standard deviation),
       and described by the Boltzmann distribution with the parameter 
       $T_{\text{eff},\,\text{K}^-}$ (c.f. eq.~(\ref{eq:teffkfi})), 
       and the normalisation factor shown in eq.~(\ref{eq:yield_kpkm}).

 \item \kam\ mesons originating from $\phi$ meson decays and observed in 
       the experiment are produced outside the collision zone.
       (Some $\phi$ mesons decay inside this zone, but the modifications of their momenta in course of the final state interactions, or the absorption effect, especially affecting \kam, may cause such events not to be reconstructable. 
       This argument implies, of course, some simplifications.)

 \item both the "direct" and the $\phi$ meson sources of \kam\ are 
       characterised by the Boltzmann distributions, with parameters: 
       $T_{\text{eff},\,\text{K}_{\text{direct}}}$ (unknown, to be found), 
       $T_{\text{eff},\,\text{K}_{\phi}}$ (c.f. eq.~(\ref{eq:teffkfi})), 
       and the normalisation factors weighting the inclusive \kam\ 
       emission probability according to the found 
       $\phi \slash \text{K}^-$ ratio (cf. eq.~(\ref{eq:fi2km})).
\end{itemize}

Fig.~\ref{fig:kmfromphi} shows the total kinetic energy distribution
of \kam\ mesons, and its two respective components.
The "direct" component was obtained by subtracting the contribution 
from the $\phi$ meson decays from the inclusive spectrum.
The Boltzmann function of the form of eq.~(\ref{eq:EkThe_Ek}) 
was fitted to this component
and the inverse slope was found to be:
%
%
%

\begin{equation}
 T_{\text{eff},\,\text{K}_{\text{direct}}}
   = 89\pm9~^{+24}_{-11}  ~\text{MeV}\quad .
\end{equation}

\noindent The uncertainties of this result take into account 
the reported statistical and systematic uncertainties of 
$P\left(\text{K}^-\right)$ and $T_{\text{eff},\,\text{K}^-}$, and of the
parameters characterising the $\phi$ meson contribution to the \kam\ spectrum.
Judging from the raw values, the inverse slope parameter of the "direct" 
antikaons (89~MeV) seems to be (somewhat) higher than the inclusive value (82~MeV), 
although not as high as the slope for the \kap\ (109~MeV). 

In a previous subsection the experimentally obtained \linebreak \kam/\kap\
ratio as function of the kinetic energy was compared to the HSD model 
calculations, in order to draw conclusions on the strength of the KN 
interaction. 
$\phi$ meson production is included in the HSD model and contributes 
to the \kam\ production. 
But the predicted yield of $\phi$'s is about 5-10 times lower 
than the measured one,
and the contribution of $\phi$ decays to the \kam\ spectra is 
assumed to be negligible for the following discussion~\cite{Leif15}.
Within the double-source model of \kam\ meson emission,
one can now substitute the "direct" \kam\ component only (instead of the
inclusive yield), as shown in fig.~\ref{fig:kmkpratio}b. 
One can find, that this subtraction results in some flattening and
lowering of the K$^-$/K$^+$ distribution profile. 
One can still uphold the statement that the calculations without the
in-medium modifications of kaonic properties does not describe the data.
However, it seems that the value of the K$^-$N potential applied 
in the HSD calculations should be somewhat lower to reproduce 
the experimental pattern.

\section{Summary}
\label{pga:sum}

FOPI has investigated the production yields and phase space distributions of 
\kap, \kam, and $\phi$ mesons from central collisions of Al+Al at a 
beam kinetic energy of 1.9A~GeV. The anisotropy parameters of the angular 
distributions, $a_2 = 0.45 \pm 0.08 ^{+0.18}_{-0.11}$ for \kap, and $a_2$ 
consistent with 0 for \kam\ were found to be in line with the findings from the 
central Ni+Ni collisions at the same beam energy. Also, the inverse slopes of kaons 
($T_\text{eff}(\text{K}^+) = 109 \pm 2 ^{+~6}_{-13}$~MeV, and
 $T_\text{eff}(\text{K}^-) =  82 \pm 6 ^{+21}_{-~6}$~MeV), 
and their total production yields per triggered event
($P_{\text{K}^+} = \left[3.75 \pm 0.07 ^{+0.33}_{-0.64} \right] \times 10^{-2}$, and
 $P_{\text{K}^-} = \left[0.95 \pm 0.10 ^{+0.26}_{-0.01} \right] \times 10^{-3}$)
were found to be consistent with the previously accumulated systematics, 
and in agreement with the well established effect of superiority of 
$T(\text{K}^+$) over $T(\text{K}^-$). 
A comparison of the measured \kam/\kap\ ratio as function of kinetic energy
to the HSD transport model calculations (a version with the negligible $\phi$ meson
production) suggests that the scenario of in-medium KN potentials
($U_{\text{K}^+\text{N}}(\rho_0) = 40$~MeV, and 
 $U_{\text{K}^-\text{N}}(\rho_0, p=0 ) \approx -50$~MeV) 
is favoured over the one without the in-medium effects. 

The emission of $\phi$ mesons was investigated in their \kap\kam\ 
decay channel. The production yield was found to be 
$\left[3.3 \pm 0.5 ^{+0.4}_{-0.8} \right] \times 10^{-4}$, 
and the inverse slope of the kinetic energy distribution
$T_{\text{eff}} = 93 \pm 14^{+17}_{-15}$~MeV. 
While the contribution of the $\phi$ decay channel into charged kaons 
to \kap\ meson emission is negligible,
the kaons originating from $\phi$-mesons were found to constitute 
$(17 \pm 3 ^{+2}_{-7})\%$ of the total \kam\ yield, a fraction 
sufficient to consider while drawing physical conclusions from 
distributions of kinematic observables involving \kam. Therefore,
the two-source model of K$^-$ emission (the "direct" component from the
collision zone, and the daughter kaons from $\phi$ decays) was applied 
to the kinetic energy spectrum of \kam. The inverse slope 
of the "direct" component was estimated to be 
$T_{\text{eff},\,\text{K}_{\text{direct}}} = 89\pm9~^{+24}_{-11}$~\text{MeV}.
The \kam/\kap\ ratio after removal of the $\phi$ contribution was found to
appear somewhat smaller and more flattened out as function of kinetic energy,
which suggests that the in-medium effects parametrised by the 
$U_{\text{K}^-\text{N}}$ potential could be weaker.

\section*{Acknowledgements}

This work was supported by the German BMBF under Contract No. 05P12VFHC7,
Korea Science and Engineering Foundation (KOSEF) under Grant No. F01-2006-000-
10035-0, by the mutual agreement between GSI and\linebreak IN2P3/CEA, by the Polish
Ministry of Science and Higher Education under Grant No. DFG/34/2007, by the National Science Centre in Poland under Grant\linebreak No. 2011/03/B/ST2/03694, by the
Hungarian OTKA under Grant No. 47168, within the Framework of the WTZ program 
(Project RUS 02/021), by DAAD\linebreak (PPP D/03/44611), and by DFG 
(Projekt\linebreak 446-KOR-113/76/04 and 436 POL 113/121/0-1).
We have also received support by the European Commission under the 6th
Framework Program under the Integrated Infrastructure on: Strongly Interacting 
Matter (Hadron Physics), Contract No. RII3-CT-2004-506078.

\bibliographystyle{elsarticle-num}
\bibliography{<your-bib-database>}

\begin{thebibliography}{}
 \bibitem{Brown91} G.E.~Brown, and M.~Rho, Phys. Rev. Lett. {\bf 66}, 2720 (1991).
 \bibitem{Weise96} W.~Weise, Nucl. Phys. A {\bf 610}, 35c (1996).
 \bibitem{Waas97} T.~Waas, M.~Rho, and W.~Weise, Nucl. Phys. A {\bf 617}, 449 (1997).
 \bibitem{Lutz04} M.F.M.~Lutz, Prog. Part. Nucl. Phys. {\bf 53}, 125 (2004), doi:10.1016/j.ppnp.2004.02.008.
 \bibitem{Fuch06} C.~Fuchs, Prog. Part. Nucl. Phys. {\bf 56}, 1 (2006), doi:10.1016/j.ppnp.2005.07.004.
 \bibitem{Hart11} C.~Hartnack, H.~Oeschler, Y.~Leifels, E.L.~Bratkovskaya,
                  and J.~Aichelin, Phys. Rep. {\bf 510}, 119 (2012), doi:10.1016/j.physrep.2011.08.004.
 \bibitem{Scha97} J.~Schaffner-Bielich, J.~Bondorf, I.~Mishustin, 
                  Nucl. Phys. A {\bf 625}, 325 (1997), doi:10.1016/S0375-9474(97)81464-9.
 \bibitem{Wisn00} K.~Wi\'sniewski {\it et al.}, Eur. Phys. J. A {\bf 9}, 515 (2000), doi:10.1007/s100500070008.
 \bibitem{Fors07} A.~F\"orster {\it et al.}, Phys. Rev. C {\bf 75}, 024906 (2007), doi:10.1103/PhysRevC.75.024906.
 \bibitem{Bena09} M.L.~Benabderrahmane {\it et al.}, Phys. Rev. Lett. {\bf 102}, 182501 (2009), doi:10.1103/PhysRevLett.102.182501.
 \bibitem{Agak10} G.~Agakishiev {\it et al.}, Phys. Rev. C {\bf 82}, 044907 (2010), doi:10.1103/PhysRevC.82.044907.
 \bibitem{Ziny14} V.~Zinyuk {\it et al.}, Phys. Rev. C {\bf 90}, 025210 (2014), doi:10.1103/PhysRevC.90.025210.
 \bibitem{Agak09} G.~Agakishiev {\it et al.}, Phys. Rev. C {\bf 80}, 025209 (2009), doi:10.1103/PhysRevC.80.025209.
 \bibitem{Lore10} M.~Lorenz, PoS (BORMIO 2010), 038.
 \bibitem{Pias15} K.~Piasecki {\it et al.}, Phys. Rev. C {\bf 91}, 054904 (2015), doi:10.1103/PhysRevC.91.054904.
 \bibitem{FOPI1}  J.~Ritman, Nucl. Phys. (Proc. Suppl.) B {\bf 44}, 708 (1995).
 \bibitem{FOPI2}  B.~Sikora, Acta Phys. Pol. B {\bf 31}, 135 (2000).
 \bibitem{Goss77} J.~Gosset {\it et al.}, Phys. Rev. C {\bf 16}, 629 (1977), doi:10.1103/PhysRevC.16.629.
 \bibitem{PDG}    K.A.~Olive {\it et al.} (Particle Data Group), 
                  Chin. Phys. C {\bf 38}, 090001 (2014), doi:10.1088/1674-1137/38/9/090001.
 \bibitem{GEANT}  wwwasdoc.web.cern.ch/wwwasdoc/geant\_html3/geantall.html
 \bibitem{IQMD}   C.~Hartnack {\it et al.}, Eur. Phys. J. A {\bf 1}, 151 (1998), doi:10.1007/s100500050045.
 \bibitem{Mang03} A.~Mangiarotti {\it et al.}, Nucl. Phys. A {\bf 714}, 89 (2003), doi:10.1016/S0375-9474(02)01366-0.
 \bibitem{Schn93} E.~Schnedermann, J.~Sollfrank, and U.~Heinz, Phys. Rev. C {\bf 48}, 2462 (1993).
 \bibitem{Lope10} X.~Lopez {\it et al.}, Phys. Rev. C {\bf 81}, 061902(R) (2010), doi:10.1103/PhysRevC.81.061902.
 \bibitem{Menz00} M.~Menzel {\it et al.}, Phys. Lett. B {\bf 495}, 26 (2000), doi:10.1016/S0370-2693(00)01232-6.
 \bibitem{Bart97} R.~Barth {\it et al.}, Phys. Rev. Lett. {\bf 78}, 4007 (1997), doi:10.1103/PhysRevLett.78.4007.
 \bibitem{Bale01} F.~Balestra {\it et al.}, Phys. Rev. C {\bf 63}, 024004 (2001), doi: 10.1103/PhysRevC.63.024004.
 \bibitem{Maed08} Y.~Maeda {\it et al.}, Phys. Rev. C {\bf 77}, 015204 (2008), doi: 10.1103/PhysRevC.77.015204.
 \bibitem{Bora05} B.~Borasoy, R.~Ni{\ss}ler, and W.~Weise, Eur. Phys. J. A {\bf 25}, 79 (2005), doi: 10.1140/epja/i2005-10079-1.
 \bibitem{HSD}    W.~Cassing {\it et al.}, Phys. Rep. {\bf 308}, 65 (1999), doi:10.1016/S0370-1573(98)00028-3.
 \bibitem{Brat03} W.~Cassing, L.~Tol\'os, E.L.~Bratkovskaya, A.~Ramos, Nucl. Phys. A {\bf 727}, 59 (2003), doi:10.1016/S0375-9474(96)00461-7.
 \bibitem{Fors03} A.~F\"orster {\it et al.}, Phys. Rev. Lett. {\bf 91}, 152301 (2003), doi: 10.1103/PhysRevLett.91.152301.
 \bibitem{RBUU1}  G.Q.~Li, G.E.~Brown, Phys. Rev. C {\bf 58}, 1698 (1998), doi:10.1103/PhysRevC.58.1698.
 \bibitem{RBUU2}  W.~Cassing {\it et al.}, Nucl. Phys. A {\bf 614}, 415 (1997), doi:10.1016/S0375-9474(96)00461-7.
 \bibitem{Hart03} C.~Hartnack, H.~Oeschler, and J.~Aichelin, Phys. Rev. Lett. {\bf 90}, 102302 (2003), doi:10.1103/PhysRevLett.90.102302.
 \bibitem{PLUTO}  I. Frohlich et al., PoS ACAT2007, 076 (2007).
 \bibitem{Leif15} Y.~Leifels (private communication).
\end{thebibliography}

%
%

\end{document}